%% file: main.tex
\documentclass[sigconf]{acmart}
\input{math_commands.tex}

\usepackage{hyperref}
\usepackage{url}
\usepackage{array}
\usepackage{booktabs}
\usepackage{subcaption}
\usepackage{pifont}
\usepackage{caption}
\usepackage{url}
\usepackage{xspace} 
\usepackage{xcolor}
\usepackage{CJKutf8}
\usepackage{multirow}
\usepackage{epsfig}
\usepackage{graphicx}
\usepackage{lscape}
\usepackage{amsmath}
\usepackage{amssymb}
\usepackage{adjustbox}
\usepackage{rotating}
\usepackage{forest}
\usepackage{amsfonts}

\newcommand{\ris}{\texttt{ReInfoSelect}}

\definecolor{midnightgreen}{rgb}{0.0, 0.29, 0.33}
\definecolor{orange}{RGB}{255,127,0}

\AtBeginDocument{%
  \providecommand\BibTeX{{%
    \normalfont B\kern-0.5em{\scshape i\kern-0.25em b}\kern-0.8em\TeX}}}

\setcopyright{acmcopyright}

\copyrightyear{2020}
\acmYear{2020} 
\setcopyright{iw3c2w3}
\acmConference[WWW '20]{Proceedings of The Web Conference 2020}{April 20--24, 2020}{Taipei, Taiwan, China} 
\acmBooktitle{Proceedings of The Web Conference 2020 (WWW '20), April 20--24, 2020, Taipei, Taiwan, China}
\acmPrice{}
\acmDOI{10.1145/3366423.3380131}
\acmISBN{978-1-4503-7023-3/20/04}

\begin{document}

\title{Selective Weak Supervision for Neural Information Retrieval}

\author{Kaitao Zhang$^\heartsuit$, Chenyan Xiong$^\spadesuit$, Zhenghao Liu$^\heartsuit$, and Zhiyuan Liu$^\heartsuit$}
\affiliation{Tsinghua University$^\heartsuit$, Microsoft Research AI$^\spadesuit$ 
} 
\affiliation{
\texttt{\{zkt18, liu-zh16\}@mails.tsinghua.edu.cn};
\texttt{chenyan.xiong@microsoft.com};
\texttt{liuzy@tsinghua.edu.cn}
}

\input{abstract.tex}


\maketitle

\input{introduction.tex}
\input{relatedwork.tex}

\input{methodology.tex}

\input{experiment.tex}
\input{evaluation.tex}

\input{conclusion.tex}

\input{acknowledge.tex}

\bibliographystyle{ACM-Reference-Format}
\normalsize
\balance
\bibliography{citation}

\end{document}

%% file: math_commands.tex
\usepackage{amsmath,amsfonts,bm}

\def\eqref#1{equation~\ref{#1}}

\def\1{\bm{1}}

\DeclareMathAlphabet{\mathsfit}{\encodingdefault}{\sfdefault}{m}{sl}
\SetMathAlphabet{\mathsfit}{bold}{\encodingdefault}{\sfdefault}{bx}{n}



%% file: abstract.tex
\begin{abstract}
This paper democratizes neural information retrieval to scenarios where large scale relevance training signals are not available.
We revisit the classic IR intuition that anchor-document relations approximate query-document relevance and propose a reinforcement weak supervision selection method, \ris{}, 
which learns to select anchor-document pairs that best weakly supervise the neural ranker (\textit{action}), using the ranking performance on a handful of relevance labels as the \textit{reward}. 
Iteratively, for a batch of anchor-document pairs, \ris{} back propagates the gradients through the neural ranker, gathers its NDCG reward, and optimizes the data selection network using policy gradients, until the neural ranker's performance peaks on target relevance metrics (\textit{convergence}).
In our experiments on three TREC benchmarks,
neural rankers trained by \ris{}, with only publicly available anchor data, significantly outperform feature-based learning to rank methods and match the effectiveness of neural rankers trained with private commercial search logs.
Our analyses show that \ris{} effectively selects weak supervision signals based on the stage of the neural ranker training, and intuitively picks anchor-document pairs similar to query-document pairs.
\end{abstract}

%% file: introduction.tex
\section{Introduction}
Neural information retrieval (Neu-IR) methods learn distributed representations of query and documents and conduct soft-matches between them in the embedding space~\cite{xiong2017knrm, convknrm, pang2017deeprank, jiafeng2016deep,macavaney2019cedr, dai2019deeper}.
In scenarios with sufficient training signals, for example, in commercial search engines with large amounts of user clicks, and on benchmarks with millions of relevance labels, end-to-end Neu-IR methods have significantly improved their ranking accuracy~\cite{convknrm, xiong2017knrm, liu2018entity, nogueira2019passage}.

Without large scale relevance labels, the effectiveness of Neu-IR is more ambivalent~\cite{yang2019critically}. 
A main challenge is that the language modeling style weak supervision, i.e. word2vec style word co-occurrence~\cite{word2vec} and BERT style mask language model~\cite{PetersELMO, devlin2019bert}, does not provide as effective distributed representations for search relevance modeling~\cite{xiong2017knrm, zhang2019generic}.
This discrepancy is a significant bottleneck for Neu-IR's impact in scenarios without the luxury of large amounts of relevance-specific supervision signals, for example, in many academic settings and none-web search domains.


This work addresses the discrepancy between weak supervision methods and the needs of relevance matching, to liberate Neu-IR from the necessity of large scale relevance-specific supervision. Inspired by the classic ``Anchor'' intuition:
\textit{anchor texts are similar to query texts and the anchor-document relations approximate relevance matches between query and documents}~\cite{croft2010search}, we propose \ris{},  ``Reinforcement Information retrieval weak supervision Selector'', which conducts selective weak supervision training specifically designed for Neu-IR models.
Given a handful of relevance labels in the target ranking task, for example, a TREC benchmark, a large amount of anchor-document pairs, and a Neu-IR model. \ris{} uses REINFORCE~\cite{williams1992simple} to learn to select anchor-document pairs that better optimize the neural ranker's performance in the target ranking task.

The weak supervision data selection is conducted by a state network that represents the anchor, the document, and the anchor-document relation, and an action network to determine whether to select each pair.
This \textit{data selector} is connected to the target \textit{neural ranker} using policy gradients---as the tool to overcome the non-differentiability of the data selection and weakly supervised training process.
The learning of the two parties is conducted iteratively in \ris{}'s stochastic process.
For a batch of anchor-document pairs, \ris{} 1) selects the weak supervision pairs using its data selector, 2) conducts several back-propagation steps through the neural ranker using the selected pairs, 3) evaluates the neural ranker on the target scenario to collect rewards, and 4) updates the data selector using policy gradients.
This process continues through batches of anchor-document pairs until the neural ranker's performance on the target task converges.

In our experiments on three widely studied TREC benchmarks, ClueWeb09-B, ClueWeb12-B13, and Robust04, \ris{} provides state-of-the-art weak supervision method for two commonly used neural rankers: Conv-KNRM~\cite{convknrm} and BERT~\cite{nogueira2019passage}.
Only when guided by \ris{}'s selective weak supervision, these neural rankers robustly outperform feature-based learning to rank methods.
\ris{} also matches the effectiveness of Bing User Clicks~\cite{dai2019deeper}---The latter is only available in commercial search environments, while everything in \ris{} is publicly available\footnote{All our codes, data, and results are available at \url{https://github.com/thunlp/ReInfoSelect}.}.

Our in-depth studies demonstrate the raw anchor-document pairs are too noisy and may hurt neural rankers' accuracy when used directly~\cite{dehghani2017neural}.
In comparison, \ris{} intuitively selects weak supervision signals based on the status of the trained neural ranker: it is lenient to noisy anchors when the neural ranker is just initialized, but, as the model converging, quickly becomes selective and only picks weak supervision signals that can further elevate the ranker's effectiveness.
This is also revealed in our human evaluation: earlier in the training process, \ris{} selects 80\%+ anchor-document pairs, most not that similar to query-relevant documents; later---as the neural ranker becomes better---\ris{} becomes more selective and picks anchor-document pairs rated as better approximations for query-relevant document pairs.

The next section discusses related work. Section 3 presents the \ris{} framework. Section 4 and Section 5 describe our experiments and results. Section 6 concludes.

%% file: relatedwork.tex
\section{Related Work}

Neu-IR models can be categorized as representation based and interaction based~\cite{jiafeng2016deep}.
Representation based methods encode the query and the document separately as two distributed representations, for example, using feed-forward neural networks~\cite{cdssm, huang2013learning}, 
convolutional networks~\cite{arcii}, or transformers~\cite{lee2019latent},
and match them using encoding distances. The encoding-and-match nature makes them more efficient for retrieval~\cite{ahmad2019reqa}.

Interaction based methods model the fine-grained interactions between query and documents, often by the translation matrices between all query and document term pairs~\cite{berger1999Information}, which in Neu-IR are calculated using term embeddings~\cite{arcii}.
The ranking scores are then calculated using the translation matrices, for example, by convolutional neural networks~\cite{arcii, Pang2016TextMA}, recurrent neural networks~\cite{pang2017deeprank}, density estimation kernels~\cite{xiong2017knrm}, and position-aware networks~\cite{hui2017pacrr}.
BERT-based rankers, with the strong interactions between query and document terms in the transformer's multi-head attentions, are also interaction based~\cite{qiao2019understanding}.

When large scale relevance training signals are available, the interaction based Neu-IR methods have shown strong effectiveness over previous feature-based learning to rank methods.
Conv-KNRM, which captures soft matches between query-document n-grams using kernels~\cite{convknrm}, significantly outperforms feature-based methods in the Chinese Sogou-T search log~\cite{zheng2018sogou, liu2018entity, zheng2019investigating}, the Bing click log~\cite{convknrm}, and MS MARCO which includes millions of expert relevance label~\cite{qiao2019understanding, hofstatter2019effect}. 
The Bing log pre-trained Conv-KNRM can also generalize to ClueWeb benchmarks~\cite{convknrm, dai2019deeper}.

Nevertheless, previous research finds that much of Neu-IR's effectiveness relies on embeddings specifically trained for relevance, because the word co-occurrence trained word embeddings do not align well with the needs of ad hoc retrieval~\cite{zamani2017relevance, xiong2017knrm}.
This necessitates large scale relevance training data for Neu-IR methods' effectiveness.
In reality, large scale relevance training data either require search logs from commercial search engines or expensive human labeling; this luxury is not often available, i.e. in academic settings or in none-web search domains where neither a large amount of human labels nor large search traffic exists. 
This significantly limits the impact of Neu-IR: in special domain search and TREC benchmarks, ambivalent performances have been observed from many Neu-IR methods~\cite{yang2019critically}.

The rise of large pre-trained transformers, e.g., BERT~\cite{devlin2019bert}, has significantly influenced the landscape of Neu-IR.
By concatenating the query and document, as a sequence to sequence pair, and fine-tuning on relevance labels, BERT based ranker outperforms previous  (shallow) neural ranking models by large margins on the MS MARCO passage ranking task~\cite{nogueira2019passage, nogueira2019document}.
The strong effectiveness, especially considering BERT is also pre-trained on word co-occurrence signals, has raised many investigations of its source of effectiveness in ranking~\cite{qiao2019understanding, dai2019deeper, padigela2019investigating}.
Though definitive conclusions remain to be studied, recent research still observed a significant defect of BERT's weak supervision in relevance matching: Dai and Callan show that when BERT is further fine-tuned on Bing user clicks, its accuracy on TREC Web Tracks improved by 16\% compared to only using Mask-LM pretraining~\cite{dai2019deeper}. This indicates that there is still a significant gap between the weak supervision of BERT's pretraining and the needs of search relevance matching.

There are several attempts to train Neu-IR models using weak supervision~\cite{luo2017training, zamani2018neural, dehghani2017learning, zamani2018theory}.
One explored weak supervision signal is Pseudo Relevance Feedback~\cite{croft2010search}.
The top retrieved documents have been used to train individual word2vec for each query and then used for query expansion~\cite{diaz2016query}.
The unsupervised retrieval scores, e.g., from BM25, have been used as relevance labels to train neural ranking models~\cite{dehghani2017neural}.
The PRF weak supervision signals in general can promote neural models to similar effectiveness as PRF-based query expansion methods~\cite{diaz2016query, dehghani2017neural, yang2019critically}.
The other explored weak supervision signal is the title-body relation in web documents, where MacAvaney et al. treat titles as an approximation of queries and build a discriminator to find most proper titles for weak supervision~\cite{macavaney2019content}.

Recently, learning to select higher quality weak supervision data has received much attention in deep learning, partly because of the ``data hungry'' property of deep neural networks~\cite{han2018co, ren2018learning, hendrycks2018using}. 
A relatively new technique, it has not yet been utilized in ad hoc retrieval nor weak supervision settings; recent studies were mainly in the domain adaptation setting of natural language processing tasks, i.e., paraphrase identification and natural language inference~\cite{qu2019learning, wang2019minimax}.

Anchor texts have been used in various IR tasks~\cite{eiron2003analysis}: query refinement~\cite{kraft2004mining}, query suggestion~\cite{dang2010query}, and document expansion~\cite{dou2009using}, to name a few. Previous usage of anchor texts in Neu-IR is mainly as an additional document field to provide additional ranking signals~\cite{zamani2018neural}, not as weak supervision. The application of reinforcement learning in IR is also mainly on ranking models~\cite{xia2017adapting, zeng2018multi}.

%% file: methodology.tex
\begin{figure*}[h]
    \centering
    \includegraphics[width=0.9\linewidth]{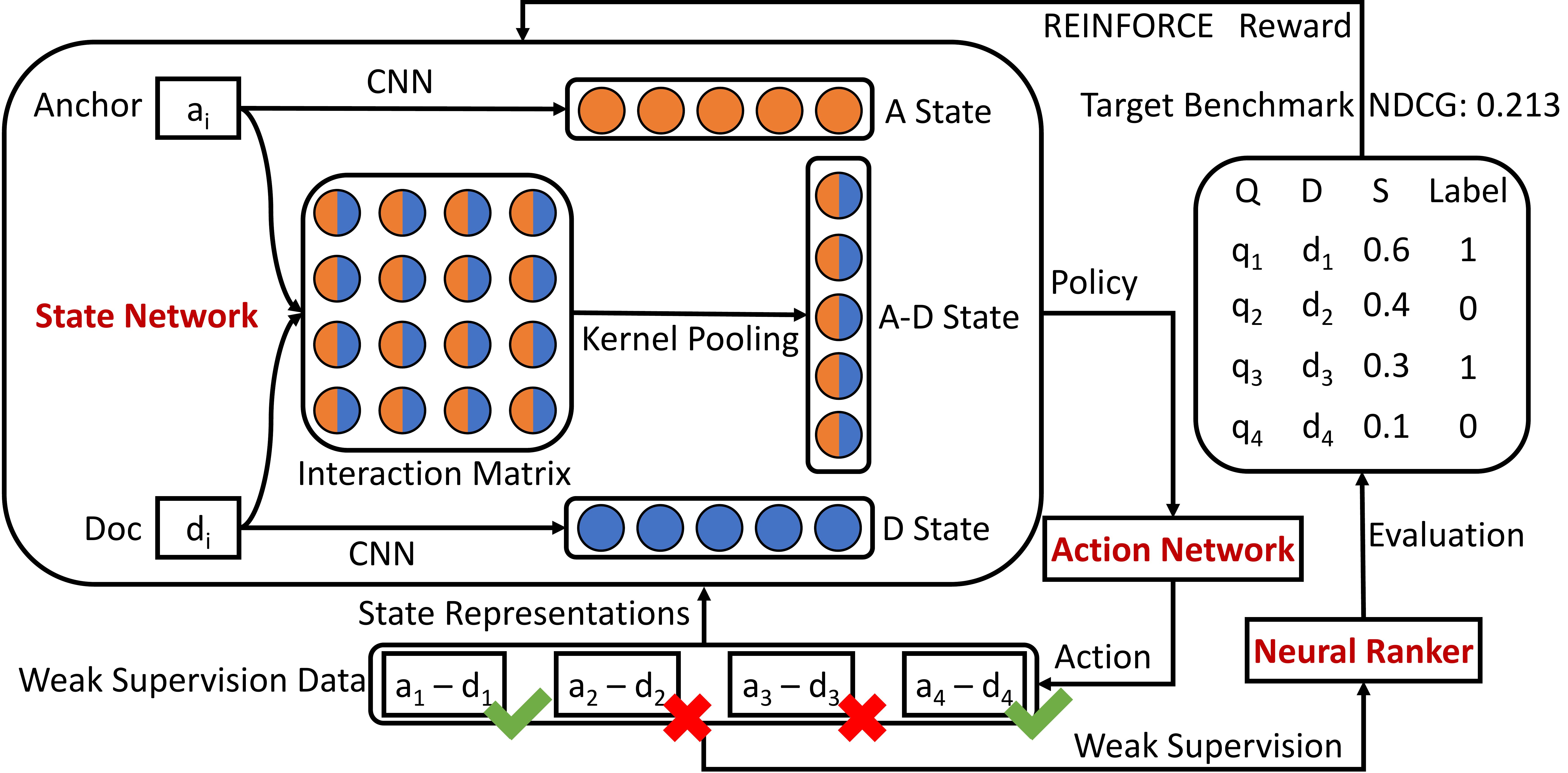}
    \caption{The Architecture of \ris{}.
    \label{fig:model}}
\end{figure*}

\section{Methodology}
This section first describes some preliminaries of neural rankers and then our \ris{} method.

\subsection{Preliminary}
Given a query $q$ and document $d$, neural ranking models calculate ranking score $f(q,d)$ using query words $q = \{ t_1^q, \dots, t_i^q, \dots, t_m^q\}$ and document words $d = \{ t_1^d, \dots, t_j^d, \dots, t_n^d \}$.
There are many neural architectures developed for $f(q, d)$. In this work, we choose Conv-KNRM~\cite{convknrm}, as the representative of neural rankers with shallow word embeddings, and BERT Ranker~\cite{nogueira2019passage}, as the representative of pre-trained transformers.

\textbf{Conv-KNRM} matches the query and document in the n-gram embedding space using matching kernels~\cite{xiong2017knrm}. It first uses Convolutional Neural Network (CNN) to calculate $h$-gram embeddings $\vec{g}_i^h$ from word embedding $\vec{t}$~\cite{convknrm}:
\begin{equation}
\vec{g}_i^h = \text{CNN}^h (\vec{t}_{i:i+h}).
\end{equation}
The $h_q$-gram and $h_d$-gram embeddings of query and document are used to construct a translation matrix $M^{h_q, h_d}$, whose item is the cosine similarity of corresponding $h$-gram pairs:
\begin{equation}
M_{ij}^{h_q, h_d} = \cos (\vec{g}_i^{h_q}, \vec{g}_j^{h_d}).
\end{equation}

It then uses $K$ Gaussian kernels to extract the matching feature $\phi(M^{h_q, h_d}) = \{ K_1(M^{h_q, h_d}), \dots ,K_K(M^{h_q, h_d})\}$ from $M^{h_q, h_d}$. Each kernel $K_k$ summarizes the translation scores as soft-TF counts:
\begin{equation}
K_k(M^{h_q, h_d}) =\sum_i (\log \sum_j \exp (- \frac{(M_{ij}^{h_q, h_d}-\mu_k)^2}{2 \delta_k^2})),
\end{equation}
where $\mu_k$ and $\delta_k$ are the mean and width for the $k$-th kernel.
The $h$-gram soft match kernels are concatenated to the final features:
\begin{equation}
\Phi(\mathcal{M}) = \phi(M^{1, 1}) \dots \circ \phi(M^{h_q, h_d}) \circ \dots \phi(M^{h_{\text{max}}, h_{d_{\text{max}}}}), \label{eq:cknrm}
\end{equation}
where $\circ$ is the concatenate operation.

The soft-TF features are combined by a standard ranking layer:
\begin{equation}
f(q,d) = \text{tanh} (\omega_r \cdot \Phi(\mathcal{M}) + b_r),
\end{equation}
with parameters $\omega_r$ and $b_r$ and $\text{tanh}$ activation.

\textbf{BERT} is a pre-trained deep transformer and performs well on many text related tasks~\cite{devlin2019bert}. 
To leverage the pre-trained BERT's sequence to sequence modeling capability, BERT first concatenates the query and document into one text sequence and feed it into pre-trained BERT~\cite{nogueira2019passage}:
\begin{align}
    \text{BERT}(q,d) = \text{Transformer}(\text{[CLS]} \circ q \circ \text{[SEP]} \circ  d \circ \text{[SEP]}).
\end{align}
The last layer's ``[CLS]'' token representation is used as the ``matching'' feature $\text{BERT}(q,d)$.
Then a ranking layer combines the representation to the ranking score:
\begin{equation}
    f(q,d) = \text{tanh} ((\omega_r \cdot \text{BERT} (q, d)) + b_r),
\end{equation}
where $\omega_r$ and $b_r$ are the learning to rank parameters.

\textbf{Training.} 
Conv-KNRM and BERT often require large scale relevance supervision~\cite{yang2019critically, guo2019deep}.
The main capacity of Conv-KNRM is its relevance specific n-gram embeddings, which need to be trained by a large amount of relevance labels, for example, user clicks in search logs~\cite{convknrm, dai2019deeper} or human labels from MS MARCO (1 Million Labels)~\cite{qiao2019understanding}.
BERT is already pre-trained by the Mask-LM task~\cite{devlin2019bert}. Still, its advantage in ranking is more observed when fine-tuned by a large amount of supervision signals from user clicks or MS MARCO~\cite{dai2019deeper, yang2019critically, nogueira2019passage, nogueira2019document}.

\subsection{Reinforcement Data Selection}
\ris{} overcomes the dependency of a large amount of relevance labels by weakly supervising neural rankers with the widely available anchor data. 
The challenge is, the anchor data are inevitably noisy: 
Many anchors are ``functional'' rather than ``informational'', for example, ``homepage'' and ``contact us''; some anchors are too general and may not even retrieve its linked document, e.g., ``customer support''. Directly using all anchor-document pairs to train Neu-IR models is unlikely effective.

To address this challenge, \ris{} learns to select more suitable anchor-document pairs directly by their ability to optimize neural rankers.
As shown in Figure~\ref{fig:model}, this is achieved by several components: the \textit{State} network that represents the a-d pair, the \textit{Action} network which decides whether to keep the pair, and the training \textit{Reward} gathered from the trained ranker.

Specifically, for the $i$-th weak supervision pair $b_i=(a_i, d_i)$, \ris{} decides if the pair should be used as a weak supervision signal (Action$_i$) by using the state representations $s_i$ of the a-d pair. The selected weak supervision pairs are then used to train the neural ranker to obtain the reward $R$, which goes back to train the action and state networks. The rest of this section describes the \textit{State}, \textit{Action}, \textit{Reward}, and the learning via \textit{Policy Gradient}.

\textbf{State.}
The state represents whether an anchor-document pair is good training data for neural rankers. The state $s_i$ for the i-th pair include three continuous vectors:
anchor state $s_i^{a}$, document state $s_i^{d}$, and anchor-document interaction state $s_i^{ad}$.

The anchor state and document state representations use standard convolutional neural networks on their word embeddings:
\begin{equation}
    s_i^{a} = \text{CNN}_a (a_i);
\end{equation}
\begin{equation}
    s_i^{d} = \text{CNN}_d (d_i).
\end{equation}
The anchor-document state representation uses the ranking feature $\Phi(\mathcal{M}_{a_i, d_i})$ from Conv-KNRM (Eqn.~\ref{eq:cknrm}):
\begin{equation}
    s_i^{ad} = \Phi(\mathcal{M}_{a_i, d_i}).
\end{equation}
The three vectors are concatenated to the final state:
\begin{equation}
    s_i = s_i^{a} \circ s_i^{d}  \circ s_i^{ad}.
\end{equation}
Note that the parameters are not shared with the neural ranker.

\textbf{Action.} 
The action decides whether to use the anchor-document pair (1) or not (0) as a weak supervision signal.
The action on the $i$-th a-d pair is calculated as
\begin{align} \label{action}
    \text{Action}_i &= \text{argmax}_{0,1} \pi(s_i), \\
    \pi(s_i) &= \text{softmax}(\text{Linear}(s_i)),
\end{align}
a simple linear layer on the state to predict the action probability.

\textbf{Reward.} The state and action networks are trained using the ultimate goal of \ris{}: the accuracy of the neural ranker when trained by the selected pairs, as the reward.

Let $\hat{B}^t = \{\hat{b}_1^t, \dots, \hat{b}_i^t, \dots, \hat{b}_{k^*}^t\}$ be a set of selected anchor-document pairs in the t-th batch $B^t$, the reward of this selected batch is:
\begin{equation}
    r^t = \text{NDCG}(f^{\hat{B}^t}(q, d)) - \text{NDCG}(f^{\hat{B}^{t-1}}(q, d)),
\end{equation}
where $f^{\hat{B}^t}$ and $f^{\hat{B}^{t-1}}$ are the neural ranker, e.g., Conv-KNRM or BERT, trained using the weak supervision pairs selected from the (1:t) and (1:t-1) batches. NDCG evaluates the neural ranker's accuracy on the \textit{validation} part of the target ranking benchmarks.
The reward is the NDCG change of the neural ranker when trained with additional weak supervision signals from $\hat{B}^t$.

\textbf{Policy Gradient.} The NDCG metric and the discrete action are not differentiable.
We use the standard policy gradient and REINFORCE to ``propagate'' the reward to the training of the state and action networks~\cite{williams1992simple}.

At the T-th reinforcement step, REINFORCE first calculates the accumulate reward $R^t$ for each of the t-th step:
\begin{equation}
    R^t = \sum_{j=t}^T c^{j} r^j.
\end{equation}
It first calculates the accumulated influence of the action taken at the t-th step in the future batches (t:T). The influence is discounted by the hyperparameter $c^{t}$. This leads to the expected reward $\overline{R}$ for the T-th step:
\begin{equation}
    \overline{R} = \frac{1}{T} \sum_{t=1}^T R^t.
\end{equation}

The expectation reward is then used to optimize the parameters of the state and action networks using standard policy gradient~\cite{williams1992simple}:
\begin{equation}
    \theta^*_{\text{state, action}} \leftarrow \theta_{\text{state, action}} + \alpha  \sum_{T} \sum_{i = 1}^{k} \overline{R} \nabla_\theta  \log \pi_\theta (s_i),
\end{equation}
where $\alpha$ is the learning rate and $k$ is the total number of weak supervision pairs of batch $B^t$. The expectation reward guides the updates of the parameters $\theta$ of the data selection policy $\pi(s_i)$.

\ris{} uses policy gradients as the tool to connect
its action and state networks to the neural ranker's performance, and thus learns to select more suitable weak supervision signals from the anchor-document pairs.

\subsection{Neural Ranker Training with \ris{}}
As shown in Figure~\ref{fig:model}, \ris{} learns interactively with the neural ranker. The two stochastically go through batches of anchor-document pairs. In each batch, \ris{} first selects the anchor-document pairs using its state and action networks, and then stochastically trains the neural ranker using the selected pairs. After that, the updated neural ranker is evaluated on the validation query-document relevance labels to obtain reward, which updates \ris{}'s policy networks via policy gradient.

Specifically, for the batch $B$, the neural ranker $f$ is trained using standard pairwise learning to rank:
\begin{equation}
l = \sum_{a_i} \sum_{d_i^{+}, d_i^{-}} \max (0, 1 - f(a_i,d_i^+) + f(a_i,d_i^{-})),
\end{equation}
where $a_i, d_i^+$ is the anchor-document pair selected by \ris{}. The document $d_i^+$ linked by $a_i$ approximates the relevant document for the pseudo query $a_i$. The negative document $d_i^-$ is from the documents retrieved by a base retrieval model, i.e., BM25, using $a_i$ as the query, following standards in learning to rank~\cite{croft2010search}.

The neural ranker is updated per batch $B$ as it is used to provide  reward for the actions taken per batch. \ris{}'s state and action networks are updated per $T$ batches (an episode in REINFORCE) to capture the action's delayed influences. The two circle through the entire anchor-document pairs and stochastically update their parameters until ranking performance converges.

%% file: experiment.tex
\section{Experimental Methodology}
This section describes the weak supervision dataset, evaluation datasets, baselines, training and implementation details.

\textbf{Weak Supervision Dataset.}
The weak supervision dataset is constructed from English corpora of ClueWeb09, which consists of 504 million web pages. The anchor texts and their linked web pages are regarded as pseudo queries and potential documents for weak supervision. All web pages are parsed by ``KeepEverythingExtractor'' in Boilerpipe. 
Anchor texts are collected using warc-clueweb\footnote{\url{https://github.com/cdegroc/warc-clueweb}}.


About $100K$ anchors (from total 6 million collected) and their linked documents are randomly sampled as the weak supervision dataset. Pseudo negative documents are the top retrieved ones by BM25 in Elastic Search\footnote{\url{https://www.elastic.co/cn/downloads/elasticsearch}}.
The influences of number of anchors and ratio of pseudo positive and negative documents are studied in Section~\ref{sec:eva:data}.

\textbf{Evaluation Datasets.} Three ad hoc retrieval benchmarks are used in evaluation: ClueWeb09-B, Robust04, and ClueWeb12-B13. 
ClueWeb09-B consists of 200 queries with relevance labels from TREC Web Track 2009-2012. Robust04 consists of 249 queries with relevance labels. ClueWeb12-B13 includes 100 queries from TREC Web Track 2013-2014. Title queries are used.

On ClueWeb09-B and Robust04, we use the exact same re-ranking setup with~\citet{dai2019deeper}, which presents the state-of-the-art neural ranking accuracy.
All our ranking models re-ranked their released top 100 SDM retrieved results~\cite{dai2019deeper}. We use the same title concatenated with the first paragraph as the document representation to fit in BERT's max sequence length~\cite{dai2019deeper}. 
On ClueWeb12-B13, we follow the re-ranking setup from~\citet{xiong2017duet}, as \citet{dai2019deeper} does not include ClueWeb12. 


All experiment settings are kept consistent with \citet{dai2019deeper}  (\citet{xiong2017duet} on ClueWeb12).
All the candidate documents to rerank are from their base retrieval methods. Conv-KNRM and BERT Ranker use open source implementations~\cite{convknrm, qiao2019understanding}.
The evaluation scores with the corresponding paper are thus directly comparable. This is crucial for reproducible IR and to compare \ris{} with the supervision from private Bing search log~\cite{dai2019deeper}.


TREC official metrics, NDCG@20 and ERR@20, are used. Statistic significance is tested by permutation test with $p<0.05$.

\textbf{Baselines.}
Our main baselines are other training methodologies in Neu-IR. We also compare with standard ranking baselines.

\texttt{No weak Supervision} uses no additional ranking labels beyond the existing small scale relevance labels in the evaluation datasets. This is the vanilla baseline. Conv-KNRM's word embeddings are initialized by Glove~\cite{pennington2014glove}; BERT uses Google's released pre-trained parameters~\cite{devlin2019bert}.

\texttt{Anchor+BM25 Labels} is our implementation of BM25 weak supervision~\cite{dehghani2017neural}. We use the same anchors in our weak supervision dataset as pseudo queries, their BM25 retrieved documents as documents, and the BM25 scores as the weak supervision labels.
The difference is that we use Conv-KNRM and BERT, stronger ranking models than their feedforward neural networks~\cite{dehghani2017neural}.

\texttt{Title Discriminator} is our implementation of MacAvaney et al.~\cite{macavaney2019content}. We use the title-document relation in ClueWeb09-B as the pseudo label and Conv-KNRM as their discriminator. It differs from \ris{} that the discriminator is not trained by reward from neural ranker's NDCG. Instead, it is a classifier trained to classify query-document pairs from title-document pairs; then the title-document pairs most similar to query-document are used.

\texttt{All Anchor} uses all randomly sampled anchor-document pairs without any filtering or weighting.

\texttt{MS MARCO Human Label} uses the passage ranking labels from MS MARCO as relevance supervision~\cite{msmarco}. It includes human labels for one million Bing queries.

\input{Tables/overall.tex}

\texttt{Bing User Clicks} is the results from Dai and Callan~\cite{dai2019deeper}, where they used user clicks in Bing as the supervision signal, which includes 5M query-document pairs. As the commercial search log is not publicly available, we use their reported numbers, which are directly comparable as our experimental setting are kept consistent.

All these training methods are applied to Conv-KNRM and BERT using the exact same setup except different training strategies.
All neural rankers are adapted to the target ranking benchmark (ClueWeb and Robust) the same with previous research~\cite{dai2019deeper, convknrm}.
They are first trained with the weak supervision or source domain supervision signals; then their ranking features (kernels or [CLS] embeddings) are combined with the base retrieval (SDM) score using Coordinate-Ascent, through standard five-fold cross validation on the target benchmark. 


Standard ranking baselines include SDM and two learning to rank methods, RankSVM and Coor-Ascent, with standard IR features. We found that the NDCG scores on ClueWeb09-B and Robust04 in~\citet{dai2019deeper} are much higher than our implementations and previous research~\cite{xiong2017knrm, convknrm}. We choose to compare with their stronger baselines, though not all metrics were provided.
ClueWeb12-B13 baselines are those released by~\citet{xiong2017duet}.

\textbf{Training Details of \ris{}.}
There are three steps for training with the \ris{}: 
warm up, reinforce training with anchor data, and adapting to the ranking benchmark.
For all steps, the ranking benchmark labels (ClueWeb09-B or Robust04) are partitioned to five folds for cross validation.

The warm up stage first trains the state and action network using the discriminator setup~\cite{macavaney2019content}. Then in the reinforce stage, \ris{}'s networks are initialized (warmed up) by the learned discriminator weights. The reinforce and adaption are all conducted via five-fold cross validation. In each of the five runs, \ris{} only uses the four training folds to calculate the rewards and to train the neural rankers. No testing information is used in any of its training stages. Then the same training splits are used to fine-tune the weakly supervised neural rankers, the same with baselines. The testing fold is only used for final evaluation.


\textbf{Implementation Details.}
This part describes the implement details of \ris{} and all baselines.

Standard Indri stopword removal and KrovetzStemmer are used to process queries and documents for Conv-KNRM. The BERT based models use BERT's sub-word tokenizer.

Conv-KNRM uses 21 kernels, one exact match and the rest soft match~\cite{qiao2019understanding}; the uni-gram, bi-gram, and tri-gram of query and document texts are considered, the same as the previous work~\cite{convknrm, dai2019deeper}. The word embedding dimension is 300 and initialed with Glove~\cite{pennington2014glove}; the learning rate is $1e-3$. The BERT models inherit pytorch-transformers\footnote{\url{https://github.com/huggingface/pytorch-transformers}}. The max sequence length is 384. Adam with learning rate $= 5e-5$ and warm up proportion 0.1 is used.

For the data selector of \ris{}, the discount factor $c^t$ is $0.99$. State networks use their own 300-dimensional word embeddings initialized with Glove~\cite{pennington2014glove}. Their CNNs use window sizes 3,4, and 5.
Its learning rate is $1e-3$.

The neural ranking models are updated with one gradient step per batch, while the data selector is updated once every 4 batches (T=4).
All our neural models are implemented with PyTorch. All models are trained with a single GeForce GTX TITAN GPU and trained about 40 hours for one epoch. More details of our implementation can be found in our code repository\footnote{\url{https://github.com/thunlp/ReInfoSelect}}.

%% file: Tables/overall.tex
\begin{table*}[t]
    \centering
    \caption{Ranking results of ReInfoSelect and baselines. $\dagger$, $\ddagger$, $\mathsection$, $\mathparagraph$, $*$ indicate statistically significant improvements over \texttt{No Weak Supervision}$^{\dagger}$, \texttt{Anchor+BM25 Labels}$^{\ddagger}$, \texttt{Title Discriminator}$^{\mathsection}$, \texttt{All Anchor}$^{\mathparagraph}$, and \texttt{MS MARCO Human Label}$^{*}$. None-neural baselines on ClueWeb09-B and Robust04 are from~\citet{dai2019deeper} and those on ClueWeb12-B13 are from~\citet{xiong2017duet}. 
    } \label{tab:overall}
    \begin{tabular}{l|l|l||l|l||l|l}
    \hline
     & \multicolumn{2}{c||}{\bf{ClueWeb09-B}} &
     \multicolumn{2}{c||}{\bf{Robust04}} & \multicolumn{2}{c}{\bf{ClueWeb12-B13}}\\ \hline
    \bf{Method} & \bf{NDCG@20} & \bf{ERR@20} & \bf{NDCG@20} & \bf{ERR@20} & \bf{NDCG@20} & \bf{ERR@20}\\ \hline
    \texttt{SDM} (From~\cite{dai2019deeper} |~\cite{xiong2017duet})
     & $0.2774$ & $0.1380$ & $0.4272$ & $0.1172$ & $0.1083$ & $0.0905$\\
    \texttt{RankSVM} (From~\cite{dai2019deeper} |~\cite{xiong2017duet})
     & $0.289$ & n.a. & $0.420$ & n.a. & $0.1205$ & $0.0924$\\
    \texttt{Coor-Ascent} (From~\cite{dai2019deeper} |~\cite{xiong2017duet})
     & $0.295$ & n.a. & $0.427$ & n.a. & $0.1206$ & $0.0947$\\
    \hline
    \textbf{Conv-KNRM as the Neural Ranker} \\ \hline
     \texttt{No Weak Supervision} (From~\cite{dai2019deeper})
     & $0.270$ & n.a. & $0.416$ & n.a. & n.a. & n.a.\\
    \texttt{No Weak Supervision} (Ours)
     & $0.2873$ & $0.1597$ & $0.4267$ & $0.1168$ & $0.1123$ & $0.0915$\\
   
    \hline
    \texttt{Anchor+BM25 Labels}~\cite{dehghani2017neural} 
     & $0.2910$ & $0.1585$ & $0.4322$ & $0.1179$ & $0.1181$ & $0.0978$\\
    \texttt{Title Discriminator}~\cite{macavaney2019content} 
     & $0.2927$ & $0.1606$ & $0.4318$ & $0.1193$ & $0.1176$ & $0.0975$\\ 
    \texttt{All Anchor}
     & $0.2839$ & $0.1464$ & $0.4305$ & $0.1190$ & $0.1119$ & $0.0906$\\ 
    \hline
    \texttt{MS MARCO Human Label} 
     & $0.2903$ & $0.1542$ & $0.4337$ & $0.1194$ & $0.1183$ & $0.0981$\\
    \texttt{Bing User Clicks} (From~\cite{dai2019deeper})
     & $0.314$ & n.a. & n.a. & n.a.
     & n.a. & n.a.\\
    \hline
    \texttt{\ris{}} 
     & $0.3094^{\dagger \ddagger \mathsection \mathparagraph *}$ & $0.1611^{\mathparagraph}$
     & $0.4423^{\dagger \ddagger \mathsection \mathparagraph *}$ & $0.1202^{\dagger}$
     & $0.1225^{\dagger \mathparagraph}$ & $0.1044^{\dagger \ddagger \mathsection \mathparagraph *}$\\
    \hline
    \textbf{BERT as the Neural Ranker} \\
    \hline
     \texttt{No Weak Supervision} (From~\cite{dai2019deeper})
     & $0.286$ & n.a. & $0.444$ & n.a. & n.a. & n.a.\\
    \texttt{No Weak Supervision} (Ours)
     & $0.2999$ & $0.1631$ & $0.4258$ & $0.1163$ & $0.1190$ & $0.0963$\\
    \hline
       \texttt{Anchor+BM25 Labels}~\cite{dehghani2017neural}
     & $0.3068$ & $0.1618$ & $0.4375^{\dagger}$ & $0.1233^{\dagger}$ & $0.1160$ & $0.0990$\\
    \texttt{Title Discriminator}~\cite{macavaney2019content}
     & $0.3021$ & $0.1513$ & $0.4379^{\dagger}$ & $0.1202^{\dagger}$ & $0.1162$ & $0.0981$\\ 
      \texttt{All Anchor} 
     & $0.3072$ & $0.1609$ & $0.4446^{\dagger}$ & $0.1206^{\dagger}$ & $0.1208$ & $0.0965$\\
     \hline
     
    \texttt{MS MARCO Human Label} 
     & $0.3085$ & $0.1652$ & $0.4415^{\dagger}$ & $0.1213^{\dagger}$ & $0.1207$ & $0.1024$\\
  
    \texttt{Bing User Clicks} (From~\cite{dai2019deeper})
     & $0.333$ & n.a. & n.a. & n.a.
     & n.a. & n.a.\\ \hline
   
    \texttt{\ris{}}
     & $0.3261^{\dagger \ddagger \mathsection \mathparagraph *}$ & $0.1669$
     & $0.4500^{\dagger \ddagger \mathsection \mathparagraph *}$ & $0.1220^{\dagger}$
     & $0.1276^{\dagger \ddagger \mathsection}$ & $0.0998$\\
    \hline
    \end{tabular}
\end{table*}

%% file: evaluation.tex
\section{Evaluation Results}
Six experiments are conducted to evaluate \ris{}'s effectiveness. We also provide human evaluations and case studies on the selected weak supervision data.

\subsection{Overall Results}

The overall ranking results are presented in Table~\ref{tab:overall}. Note that though our implementation of No Weak Supervision performs slightly better than Dai and Callan's~\cite{dai2019deeper}, with only the several hundreds of labeled queries, neither Conv-KNRM nor BERT convincingly outperforms the classic feature-based learning to rank methods, \texttt{RankSVM} or \texttt{Coor-Ascent}. Similar ambivalent effectiveness on Neu-IR systems when relevance training data are limited has been observed in multiple previous studies~\cite{yang2019critically, dai2019deeper, convknrm, xiong2018towards}.

With both neural ranking models, \ris{} outperforms all baselines except \texttt{Bing User Clicks} on both datasets. The improvements on NDCG are robust across the table, while the ERR metric is a little more brittle, especially on ClueWeb12-B13, the same as observed by previous research~\cite{convknrm, macavaney2019cedr, xiong2018towards}. 
\ris{} and \texttt{Bing User Clicks} are the only two methods that show stable improvements over the feature-based \texttt{Coor-Ascent}. Note again that \texttt{Bing User Clicks} are not publicly available, while \ris{} only uses widely available anchor information.

Adapting from \texttt{MS MARCO Human Label} does not lead to much improvement, though using one million expert relevance labels. The MS MARCO ranking task is a passage ranking for more natural language queries. The domain differences limit the generalization ability of human relevance labels. Weak supervision and transfer learning sometimes are necessary: One million human labels are not time and cost effective, if ever feasible, in many ranking scenarios.

\ris{} significantly outperforms \texttt{All Anchor} on both datasets. The latter uses the same anchor-document relation as weak supervision signals but without any selection. Random anchors are noisy and not always similar to search queries. Section~\ref{sec:selectedanchor} further studies the quality of selected anchors.

Our implementation of \texttt{Anchor+BM25 labels} uses better neural rankers and also combines it with base retrieval, compared to its vanilla form in previous research~\cite{dehghani2017neural}. Still, it does not yet outperform \texttt{No Weak Supervision}. BM25 scores can be used as pseudo relevance feedback (PRF) or to find stronger negative documents~\cite{croft2010search, macavaney2019content}.
More in line with the later,  \ris{} uses BM25 to find negative documents for anchors. 

There are two differences between \texttt{Title} \texttt{Discriminator}~\cite{macavaney2019content} and \ris{}. The first is the weak supervision signals: Title VS Anchor. The second is that \ris{} learns to select weak supervision pairs using reward from the neural ranker, while \texttt{Title Discriminator} conducts the selection of training data and the training of neural rankers independently~\cite{macavaney2019content}. Section~\ref{sec:disc} and Section~\ref{sec:select} further study the effectiveness of the data selection.

We have also experimented with other neural rankers, including EDRM, which integrates external knowledge~\cite{liu2018entity}, and Transformer-Kernel, which uses sub-word and transformer with kernels~\footnote{https://github.com/sebastian-hofstaetter/transformer-kernel-ranking}. Similar effectiveness and trends were observed.
We also explored the ensemble of the \ris{} supervised neural rankers; similar gains from previous research are observed~\cite{pyreddy2018consistency} and the ensemble models outperform the single model trained by Bing user Clicks, though it is not a fair comparison. These additional results are listed in our open source repository due to space limitations.

\subsection{State Networks As Discriminators}
\label{sec:disc}





This experiment studies the effectiveness of the state representations, using an intermediate evaluation---how well they can classify the actual query-documents from anchor-documents pairs, i.e., as the discriminator role in \texttt{Title} \texttt{Discriminator}~\cite{macavaney2019content}.

We use the state network and the action network to directly learn a binary classifier, using the query-documents from ClueWeb09-B, Robust04 or ClueWeb12-B13 as positive instances and the anchor-documents as negative instances. We train the model the same as we train the \texttt{Title} \texttt{Discriminator} and evaluate their classification accuracy in five-fold cross-validation. The results are shown in Table~\ref{tab:policy:state}.

Overall, the ClueWeb queries and documents are harder to distinguish from the anchor and documents; the accuracy on ClueWeb in general is lower than on Robust. This is expected as the weak supervision documents are also from ClueWeb and the ClueWeb queries are in the web domain. There is less domain difference between our weak supervision and ClueWeb's labels. This also correlates with the greater relative improvements of \ris{} on ClueWeb compared to Robust.

All three state networks have decent accuracy on both datasets, showing their good representation ability of the query/anchor-document pairs. The Anchor State easily distinguishes anchors from queries, and the Anchor-Document states find query-document pairs from anchor-document pairs, especially on Robust.
However, these intermediate results only show whether the state networks represent the data well. It is unclear, especially when used in the discriminator setup~\cite{macavaney2019content}, whether a highly accurate classifier will lead to better trained neural models. A perfect discriminator discards all anchor-document data, which is not useful as it leaves no information to use for weak supervision. We care the most whether these state networks help us pick weak supervision pairs that lead to more effective neural rankers, which is studied in the next experiment.

\input{Tables/policy.tex}

\subsection{Effectiveness of \ris{} Selection} 
\label{sec:select}

\input{Figures/reinforce.tex}

\input{Tables/finetune.tex}

 Table~\ref{tab:policy:class} shows the effectiveness of \ris{}'s data selection with different strategies. 
 Three different policy strategies are experimented.
 The first is the two-step approach, \texttt{Discriminator}~\cite{macavaney2019content}. It selects anchor-document pairs that like real query-document pairs and then train the neural ranker with selected data. 
 The next two are from \ris{}:  \texttt{Scratch} initializes its state and action network from scratch, and \texttt{Warm Up} initializes its state and action network from the results of the corresponding \texttt{Discriminator}.
 Three state combinations are evaluated with the three strategies: A(nchor) State only, A(nchor)-D(ocument) State only, and All State. 

All data selection methods outperform the \texttt{ALL Anchor}, which illustrates the anchor data is informative but noisy; data selection is necessary to filer out noisy anchor data. The reinforce based data selection methods outperform all \texttt{Discriminator} models, demonstrating the effectiveness of \ris{}. Connecting the performances of weakly supervised neural ranker to data selector using the policy network provides significant accuracy boosts.

\texttt{All state} shows better performance than other state representations. It helps \ris{} select better anchor-document pairs for weak supervision, though it might not be the best classifier to distinguish anchor from the query when used as a data discriminator. 
We also find \texttt{Warming Up} the state and action networks slightly more effective than training from \texttt{Scratch}. We observe that the REINFORCE is slow and unstable in training and a better initial state may improve its stability. This is further studied in Section~\ref{sec:curves}.

\subsection{Fine-Tuning Strategies}
\input{Figures/datastrategy.tex}

\begin{figure*}[t]
    \centering
        \begin{subfigure}[b]{0.23\textwidth}
        \includegraphics[width=\textwidth]{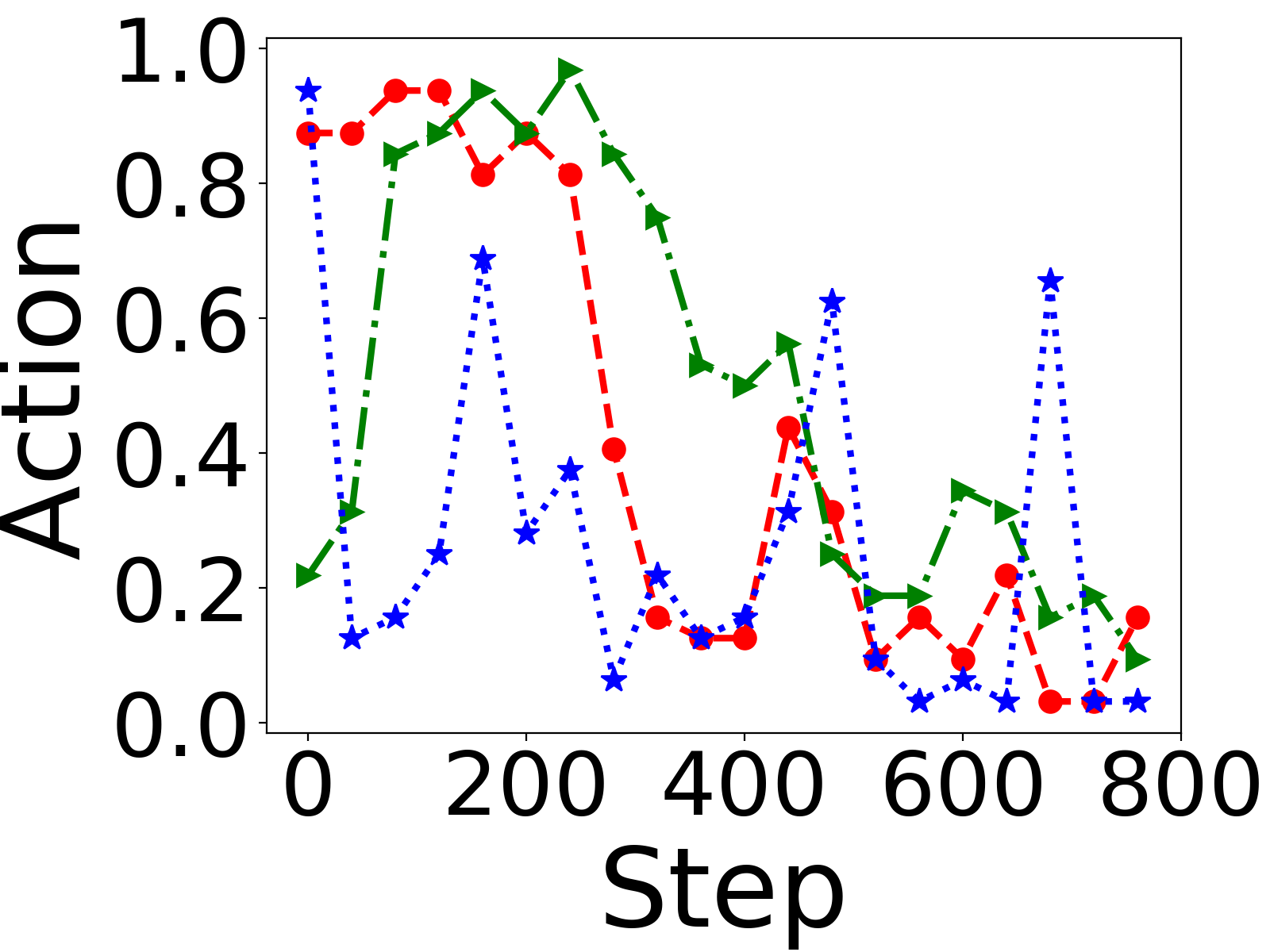}
        \centering
        \caption{Action w. Conv-KNRM}
    \end{subfigure}
    \begin{subfigure}[b]{0.23\textwidth}
        \includegraphics[width=\textwidth]{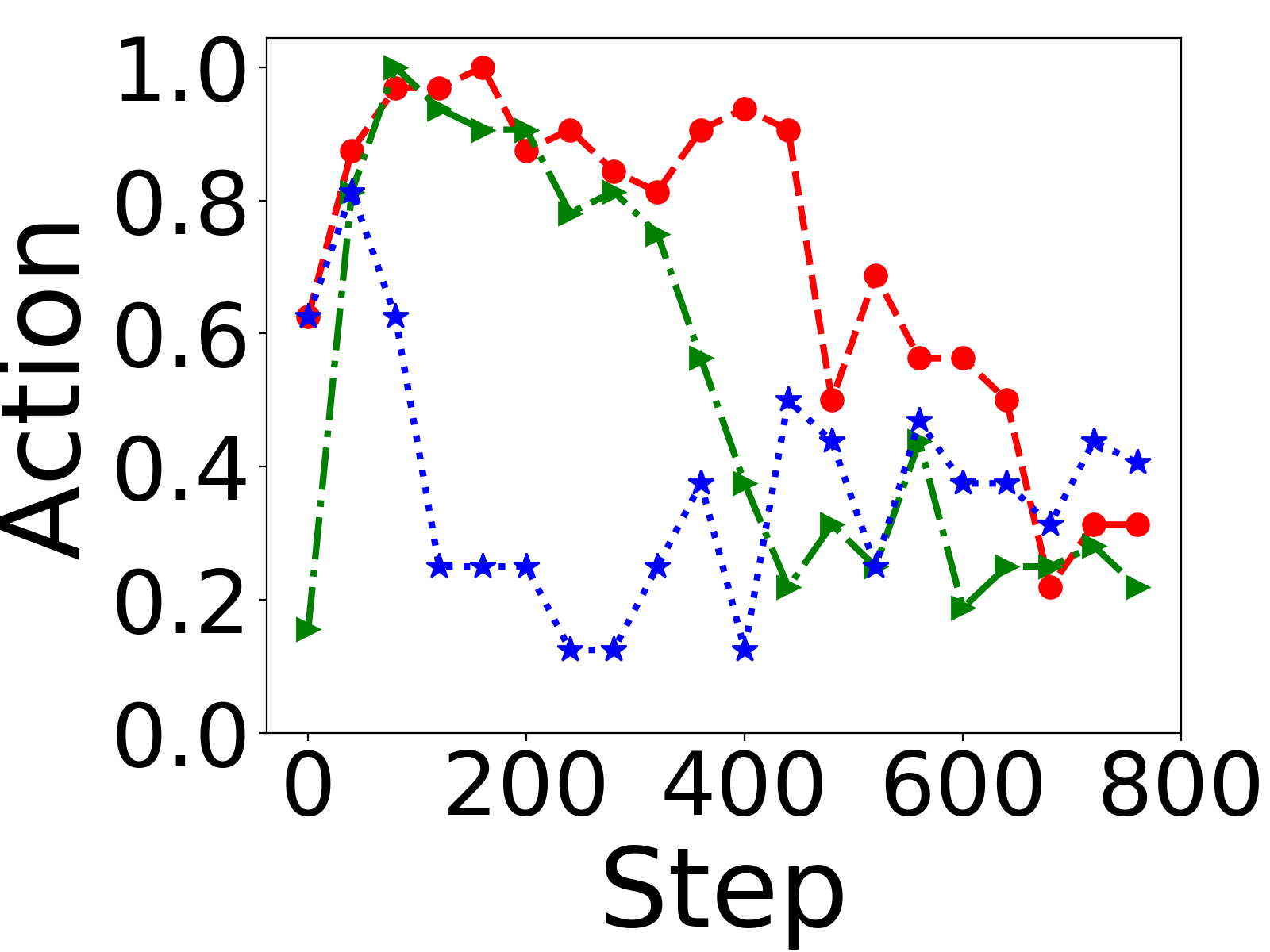}
        \centering
        \caption{Action w. BERT}
    \end{subfigure}
        \begin{subfigure}[b]{0.23\textwidth}
        \includegraphics[width=\textwidth]{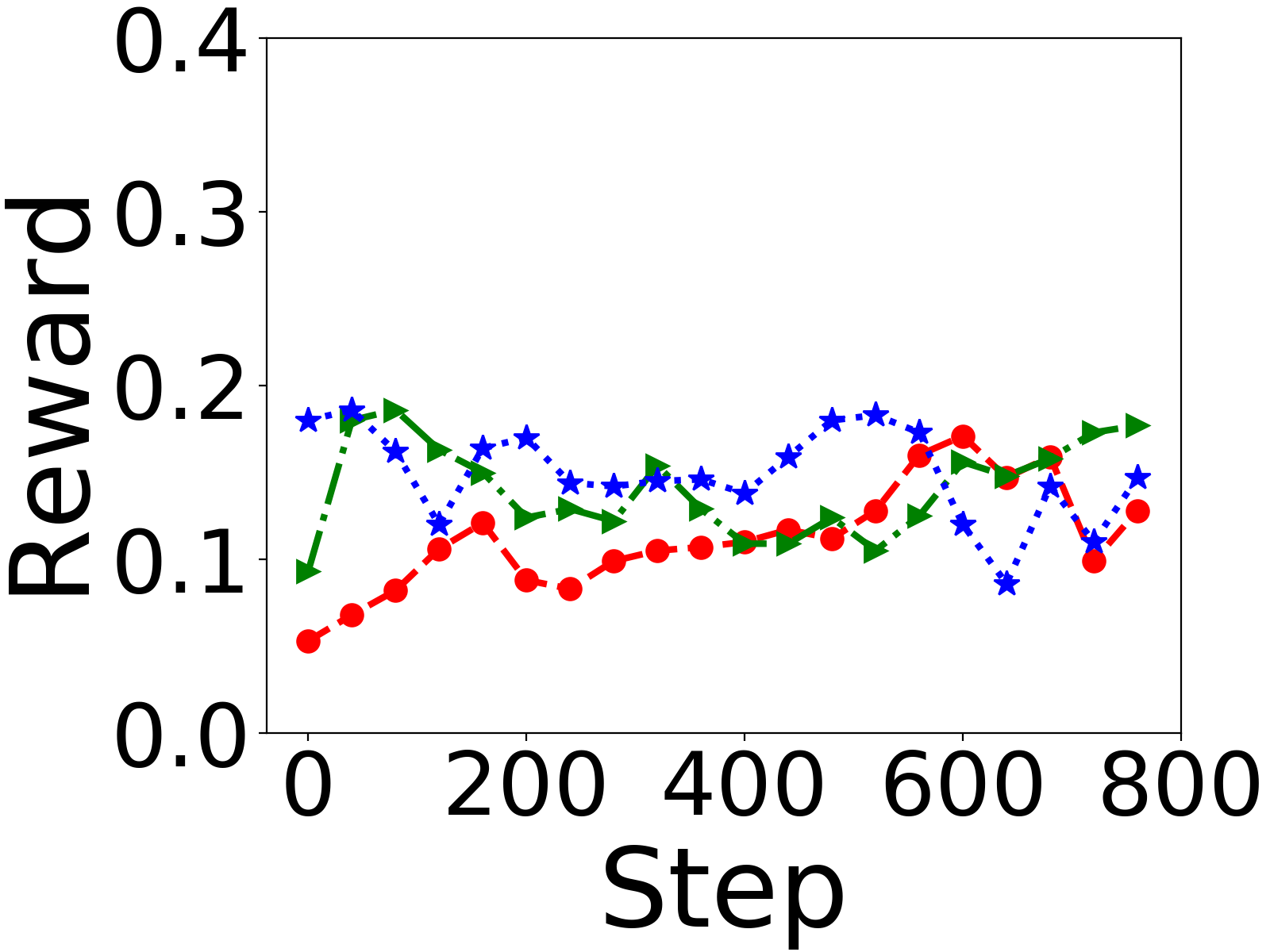}
        \centering
        \caption{Conv-KNRM Reward}
    \end{subfigure}
    \begin{subfigure}[b]{0.23\textwidth}
        \includegraphics[width=\textwidth]{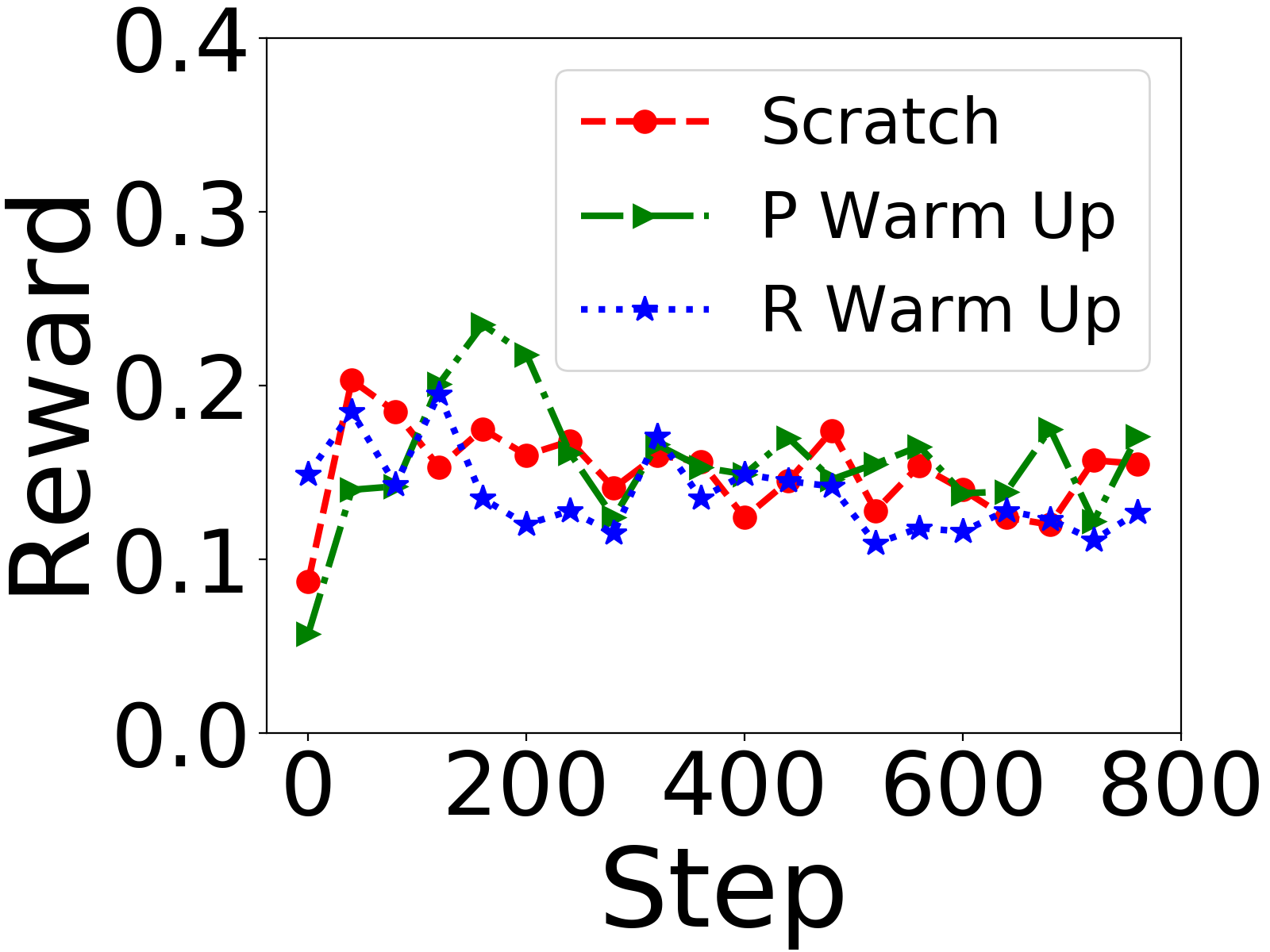}
        \centering
        \caption{BERT Reward}
    \end{subfigure}
    \caption{The behavior of \ris{}'s on ClueWeb09-B when trained from \texttt{Scratch}, with its policy networks warmed up (\texttt{P Warm Up}), and the neural ranker warmed up (\texttt{R Warm Up}), both from \texttt{All Discriminator}. 
    X-axes show the training step (batches) before convergence; Y-axes show the fraction of pairs being selected (action) and the rewards from neural rankers.
    \label{fig:training}}
    
\end{figure*}

This experiment studies different fine-tuning strategies when adapting the weakly supervised neural rankers to target ranking scenarios. We focus on Conv-KNRM as the ranker and ClueWeb09-B as the target ranking scenario, and experiment with several different adaption approaches. The first directly fine-tunes the \texttt{Dense} \texttt{Layer} of Conv-KNRM on ClueWeb09-B, with the embeddings frozen (which we find more effective). The second feeds the kernel scores (soft match features) to \texttt{Coor-Ascent} instead of a Dense layer. The last is the one used in previous research and this work, which combines the neural ranker's features with SDM scores in the standard learning to rank set up~\cite{convknrm, dai2019deeper}. We also combine Conv-KNRM and BERT with SDM, following previous research~\cite{macavaney2019cedr}.

Adding \texttt{SDM} score significantly improves the ranking accuracy. Even Conv-KNRM has the soft n-gram match function. The core IR intuitions, e.g., proximity, smoothing, and normalization, in SDM are still necessary for neural rankers to perform well. The current Neu-IR models enhance classic IR approaches but have not yet replaced them. 
\texttt{Coor-Ascent} is also much more effective learning to rank model compared to the simple Dense (Linear) Layer. The listwise ranker is as effective in combining neural features as in combining classic features.

Combing Conv-KNRM and BERT also provides further improvements over either one individually~\cite{macavaney2019cedr}. We also observed better performances of \ris{} over \texttt{Discriminator} across all settings, showing the robust effectiveness of \ris{} in selecting more effective weak supervision signals.

\subsection{Influence of Training Data Strategies}
\label{sec:eva:data}

Recent advancements of Neu-IR models are mainly trained on data with a vast amount of queries but only a handful of document labels per query, e.g. search log clicks~\cite{convknrm, dai2019deeper} and MS MARCO labels~\cite{nogueira2019document}. Both have less than five relevance documents per query, much fewer than typical TREC benchmarks.

This experiment studies the influence of different data combinations in training neural ranking models. Specifically, we keep the number of total anchor-document pairs roughly the same and vary the different combinations of the number of anchors (\#a) and documents per anchor (\#d/a). We also try some different balances of positive VS negative documents. The results of \ris{} trained Conv-KNRM on ClueWeb09-B are shown in Table~\ref{tab:data}.

The ranking accuracy does vary, to some degree, with different data combinations, although using similar amounts of training labels. We observe that Conv-KNRM prefers more variations on the query side. It performs better when with more anchors but fewer documents per anchor. This correlates with the setup in MS MARCO and in Sogou-QCL~\cite{msmarco, zheng2018sogou}. We also observe that the neural ranker prefers more positive labels per query. It is expected as the positive labels provide more information than negative ones. Nonetheless, in this experiment, \ris{} is not very sensitive and its accuracy does not vary much across different data strategies.

\subsection{Stochastic Training Analysis}
\label{sec:curves}

This experiment analyzes the stochastic training of \ris{}. We plot the action (fraction of selected pairs) and the reward of each training batch in Figure~\ref{fig:training}. 
Three variations of \ris{} are evaluated: with all parameters in the policy network and the neural rankers from \texttt{Scratch} (BERT ranker is initialized from pretraining without IR continuous training), with the policy network parameters warmed up using parameters from \texttt{All Discriminator}, and with the neural rankers warmed up using their correspondence from \texttt{All Discriminator}.


\texttt{Scratch} starts by selecting a large fraction of weak supervision pairs, and quickly boosts the reward from the neural rankers. At the beginning of \ris{} training, the neural rankers are nearly random, and raw anchor-document pairs can help boosts their performance. As the training goes on, the neural rankers get better and better, thus being more selective in their training data. \ris{} then discards more of the anchor-document pairs as adding them may lead to negative rewards. 

The same intuitions are reflected by the two warm up versions as well.
In \texttt{P Warm Up}, the model starts with a low selection rate, as the policy network is initialized by \texttt{All Discriminator}, which is trained to consider all anchors as negatives. However, it quickly learns to relax the selection rate, as the neural ranker is just initialized and most anchor-document pairs can help. After that, \texttt{P Warm Up} behaves rather similarly to \texttt{Scratch}.
In \texttt{R Warm Up}, as the neural ranker is warmed up with meaningful weights, the data selection rates remain low and \ris{} only picks those can further boosted the warmed up ranker's accuracy.

The behaviors of \ris{} with Conv-KNRM and BERT share similar trends in the stochastic training process.
The main difference is that BERT uses more relaxed selection rate for more epochs. Its deep transformer networks are hard to optimize than Conv-KNRM's embedding and CNN layers. 
On the reward side, all model variations converge to similar reward scores, showing the robustness of \ris{}.


\subsection{Data Selection Behaviors}
\label{sec:selectedanchor}
\input{Figures/agreement.tex}

This set of experiments analyze the data selection behaviors of \ris{}, include the agreements between different models and human evaluations.

\textbf{Method Agreements.} The first experiment studies whether \ris{} chooses different weak supervision signals for different neural rankers.
Table~\ref{tab:agreement} shows the agreement between the ``All \texttt{Discriminator}'' and \ris{} when used with the two neural rankers. The agreements between the same model (diagonal elements) are evaluated on the five cross-validation runs. The rests are the average of between their five runs. All models read the anchor-document pairs in the same order.

\input{Figures/human.tex}

All methods agree the most with themselves across different runs. \texttt{Discriminator} is more deterministic than \ris{}, as expected.
\ris{} picks different data for Conv-KNRM and for BERT. The latter has already been pre-trained and has rather different, and much deeper, architectures than Conv-KNRM. 
\ris{} disagrees with \texttt{Discriminator} the most. The data selection in \texttt{Discriminator} is disconnected from the target ranker.

\input{Tables/case_dis.tex}
\input{Tables/case.tex}

\textbf{Human Evaluation.} This experiment examines the anchor intuition and evaluates whether \ris{} selects anchors that are considered reasonable web queries and whether the documents are considered as relevant for their anchors.

We recruit four graduate students to label anchor-document pairs. We present them 100 anchors, each associated with one linked document (pseudo positive) and one retrieved (pseudo negative) document, all randomly sampled and ordered. The judges are asked to provide two binary labels. The first is whether the anchor could be a reasonable web search query, and the second is which of the two documents is more relevant to the anchor. The majority of votes are used as the final label. Our judges agree well: their Cohen's Kappa is $0.526$ on anchor and $0.544$ on anchor-document.

We then mix the labeled pairs into all pairs and feed them to \texttt{Discriminator} and \ris{}. When feeding \ris{}, we either mix the labeled pairs in the early part of the epoch, when the neural ranker is just initialized, and the late part, when the neural ranker is close to converge, to evaluate the behavior on the two stages. The results are shown in Table~\ref{tab:human}. 

Aligned with the Anchor intuition, 90\%+ anchors are rated as reasonable search queries. The relevance between anchors and linked documents is more ambivalent; many anchors are functional (e.g. ``homepage'') than informational. 
\ris{} shows significantly different behavior in Early and Late. As the neural ranker converging, \ris{} selects pairs more and more similar to query-relevant documents, while in the beginning, especially with the less pre-trained Conv-KNRM, 91\% data is selected. Its ``selectiveness'' is customized for the target's training status.

\texttt{Discriminator} does a better job in picking a-d pairs similar to query and relevant documents, which is what it is trained for.
However, it is too strict and only picks 5\% pairs.
The challenge is that it is only trained with limited target query and documents, thus may not generalize well to anchors that are good but different from the handful target queries. Another challenge is that the data selector is isolated with the target ranker, while \ris{} provides some final push that elevates the target neural rankers another 3-5\% compared to \texttt{Discriminator}.

\subsection{Case Study}
We first study instances where \ris{} and \texttt{Discriminator} chose different actions. The example anchor-document pairs are listed in Table~\ref{tab:case:anchor:doc}. 
Some anchor and document pairs are quite similar to search queries and relevant documents, for example, ``kansas city zoo'' leads to the Yahoo! page about it. There are anchors that are more functional, e.g., ``banner ads'', and too general, e.g., ``power load''. The linked documents may also be irrelevant. For example, the pictures published by New York State Police are as irrelevant to ``New York State Police''.  \texttt{Discriminator} and \ris{} behave differently and as shown in last experiment, the difference was mainly from the (dis)connection with the target Neu-IR model.

We also find many anchors selected by \ris{} very similar to the actual queries in the TREC benchmarks. Table~\ref{tab:case:anchor} lists some examples. These anchors reflect very similar information needs with the actual queries.
For example, ``vegan menu for people with diabetes'' is a legit web search query; one can imagine it forming a search session with ``diabetes education''. These echo the classic IR intuition that many anchors are similar to search queries, which are effectively selected by \ris{}.

%% file: Tables/policy.tex
\begin{table}[]
    \centering
    \caption{Classification Accuracy of different state networks when used as the data discriminator~\cite{macavaney2019cedr}: \texttt{A}nchor only, \texttt{A}nchor-\texttt{D}ocument pair only, and \texttt{All} state together.}\label{tab:policy:state}
    \small
    \begin{tabular}{l|c|c||c|c||c|c}
    \hline
    \bf{Method} & \multicolumn{2}{c||}{\bf{ClueWeb09-B}} & \multicolumn{2}{c||}{\bf{Robust04}} & \multicolumn{2}{c}{\bf{ClueWeb12-B13}} \\
    \hline
    \bf{Discriminator} & \bf{A} & \bf{A-D} & \bf{A} & \bf{A-D} & \bf{A} & \bf{A-D}\\ \hline
    \texttt{A State} & $0.775$ & -- & $0.850$ & -- & $0.786$ & -- \\
    \texttt{A-D State} & $0.723$ & $0.702$ & $0.890$ & $0.928$ & $0.726$ & $0.712$ \\
    \texttt{All State} & $0.743$ & $0.728$ & $0.863$ & $0.899$ & $0.740$ & $0.732$ \\
    \hline
    \end{tabular}
\end{table}

%% file: Figures/reinforce.tex
\begin{table}[t]
    \centering
    \caption{Conv-KNRM Results on ClueWeb09-B in different classifiers or \ris{} states. Relative performances \% and statistical significance$^\dagger$ are compared with the corresponding Discriminator using the same states.\label{tab:policy:class}}
    \small
    \begin{tabular}{l|lr|lr}
    \hline
    \bf{Method} & \multicolumn{2}{c|}{\bf{NDCG@20}} & \multicolumn{2}{c}{\bf{ERR@20}}\\
    \hline
    \texttt{All Anchor} & $0.2839$ & -- & $0.1464$ & -- \\
    \hline
     \texttt{A Discriminator}  & $0.2893$ & -- & $0.1521$ & -- \\
    \texttt{A State (Scratch)} & ${0.3002}^{\dagger}$ & $+3.77\%$ & $0.1604$ & $+5.46\%$ \\
    \texttt{A State (Warm Up)} & ${0.3050}^{\dagger}$ & $+5.43\%$ & $0.1632$ & $+7.30\%$ \\ \hline
    
    \texttt{A-D Discriminator} & $0.2974$ & -- & $0.1556$ & -- \\
    \texttt{A-D State (Scratch)} & $0.3033$ & $+1.98\%$ & $0.1653$ & $+6.23\%$ \\
     \texttt{A-D State (Warm Up)} & ${0.3083}^{\dagger}$ & $+3.66\%$ & $0.1646$ & $+5.78\%$ \\
    \hline
    
     \texttt{All Discriminator} & $0.3021$ & -- & $0.1576$ & -- \\ 
         \texttt{All State (Scratch)} & $0.3078$ & $+1.89\%$ & $0.1670$ & $+5.96\%$ \\
    \texttt{All State (Warm Up)} & ${0.3094}^{\dagger}$ & $+2.42\%$ & $0.1611$ & $+2.22\%$ \\
    \hline
    \end{tabular}
\end{table}

%% file: Tables/finetune.tex
\begin{table*}[]
    \centering
    \caption{\texttt{Conv-KNRM} Results on ClueWeb09-B in different fine-tuning strategies. Percentages indicate relative performance over \texttt{No Weak Supervision}.\label{tab:finetune}}
    \begin{tabular}{l|l|l|lr|lr}
    \hline
    \bf{Method} & \bf{Feature}& \bf{LeToR}& \multicolumn{2}{c|}{\bf{NDCG@20}} & \multicolumn{2}{c}{\bf{ERR@20}}\\ \hline
    \texttt{No Weak Supervision (Ours)} & \texttt{Conv-KNRM+SDM} & \texttt{Coor-Ascent} & $0.2873$ & -- & $0.1597$ & -- \\
    \hline
    \texttt{Discriminator} & \texttt{Conv-KNRM} & \texttt{Dense Layer} & $0.2523$ & $-12.18\%$ & $0.1344$ & $-15.84\%$ \\
     & \texttt{Conv-KNRM} & \texttt{Coor-Ascent} & $0.2787$ & $-2.99\%$ & $0.1429$ & $-10.52\%$ \\
     & \texttt{Conv-KNRM+SDM} & \texttt{Coor-Ascent} & $0.2980$ & $+3.72\%$ & $0.1592$ & $-0.31\%$ \\
     & \texttt{Conv-KNRM+BERT+SDM} & \texttt{Coor-Ascent} & $0.3170$ & $+10.34\%$ & $0.1747$ & $+9.39\%$ \\
    \hline
    \texttt{ReInfoSelect} & \texttt{Conv-KNRM} & \texttt{Dense Layer} & $0.2694$ & $-6.23\%$ & $0.1523$ & $-4.63\%$ \\
     & \texttt{Conv-KNRM} & \texttt{Coor-Ascent} & $0.2896$ & $+0.80\%$ & $0.1615$ & $+1.13\%$ \\
     & \texttt{Conv-KNRM+SDM} & \texttt{Coor-Ascent} & $0.3094$ & $+7.69\%$ & $0.1611$ & $+0.88\%$ \\
     & \texttt{Conv-KNRM+BERT+SDM} & \texttt{Coor-Ascent} & $0.3222$ & $+12.15\%$ & $0.1796$ & $+12.46\%$ \\
    \hline
    \end{tabular}
\end{table*}

%% file: Figures/datastrategy.tex
\begin{table}[]
    \centering
    \caption{Data Strategies Results of ReInfoSelect on ClueWeb09-B, with different number of anchors (\#a), different number of positive/negative documents per anchor (\#d+/a and \#d-/a), and different number of total anchor-document pairs (\#pair).
    Percentages indicate relative performance compared with \texttt{10K} anchors.
    }\label{tab:data}
    \small
    \begin{tabular}{r|r|r|r|lr|lr}
    \hline
    \bf{\#a}& \bf{\#d$^+$/a} & \bf{\#d$^-$/a} & \bf{\#pair} & \multicolumn{2}{c|}{\bf{NDCG@20}} & \multicolumn{2}{c}{\bf{ERR@20}}\\
    \hline
      \texttt{10K} & $35.4$ & $35.5$ & $0.7M$ & $0.3025$ & -- & $0.1611$ & --\\
      \texttt{50K} & 7.2 & $7.2$ & $0.7M$ & $0.3084$ & $+1.95\%$ & $0.1651$ & $+2.48\%$ \\
      \texttt{100K} & $5.5$ & $5.5$ & $1.1M$ & $0.3094$ & $+2.28\%$ & $0.1611$ & $+0.00\%$ \\
      \texttt{100K} & $3.6$ & $7.1$ & $1.1M$ & $0.3073$ & $+1.59\%$ & $0.1645$ & $+2.11\%$ \\
      \texttt{100K} & $2.0$ & $8.0$ & $1.0M$ & $0.3042$ & $+0.56\%$ & $0.1621$ & $+0.62\%$ \\
      \texttt{500K} & $1.2$ & $1.2$ & $1.2M$ & $0.3085$ & $+1.98\%$ & $0.1664$ & $+3.29\%$
      \\
    \hline
    \end{tabular}
\end{table}

%% file: Figures/agreement.tex
\begin{table}[t]
    \centering
    \caption{Method Agreements on ClueWeb09-B: the fraction of anchor-document pairs where different runs/methods choosing the same action (select/not select). $w/$ represents \ris{} uses the following model as neural ranker.
    }\label{tab:agreement}
    \small
    \begin{tabular}{l|c|c|c}
    \hline
    \bf{Method} & \texttt{Discriminator} & \texttt{w/ Conv-KNRM} & \texttt{w/ BERT} \\
    \hline
    \texttt{Discriminator} & \textbf{0.924} & $0.356$ & $0.109$ \\ \hline
    \texttt{w/ Conv-KNRM} & $0.356$ & \textbf{0.585} & $0.323$ \\
    \texttt{w/ BERT} & $0.109$ & $0.323$ & \textbf{0.623} \\
    \hline
    \end{tabular}
\end{table}


%% file: Figures/human.tex
\begin{table}[t]
    \centering
    \caption{Human Evaluation on ClueWeb09-B. 
    The numbers are the fraction of anchors (A) or anchor-document pairs (A-D) labeled as proper search queries or relevant query-document pairs. 
    Selected and Discarded are the actions taken by the models. Percentages ($\%$) are the fraction of Selected (the same on A and A-D).
    Early and Late refer to the training stage in \ris{}: before and after 400 batches. 
    }\label{tab:human}
    \small
    \begin{tabular}{l|c|c|c|c|c}
    \hline
    & \multicolumn{2}{c|}{\textbf{Selected}} & \multicolumn{2}{c|}{\textbf{Discarded}} & \textbf{Selected \%} \\ \hline
    & \textbf{A} & \textbf{A-D} & \textbf{A} & \textbf{A-D} & \textbf{A \& A-D} \\ \hline
    \texttt{Discriminator} & 0.900 & 0.750 & 0.923 & 0.479 & 5\% \\ \hline 
    \textbf{With Conv-KNRM}  \\ \hline
    \texttt{\ris{}-Early} & 0.926 & 0.489 & 0.861 & 0.528 & 91\% \\ 
    \texttt{\ris{}-Late} & 0.942 & 0.540 & 0.892 & 0.432 & 56\% \\ \hline
    \textbf{With BERT}  \\ \hline
    \texttt{\ris{}-Early} & 0.912 & 0.510 & 0.946 & 0.435 & 77\% \\ 
    \texttt{\ris{}-Late} & 0.929 & 0.560 & 0.912 & 0.378 & 63\% \\ \hline
    \end{tabular}
\end{table}

%% file: Tables/case_dis.tex
\begin{table*}[t]
    \centering
    \caption{Case study for anchor-document pairs that only selected by \texttt{Discriminator} or \ris{}. Document snippets are manually picked. Manual labels on whether the anchor or pair is relevant (+) or not (-) are shown in brackets. }\label{tab:case:anchor:doc}
    \small
    \begin{tabular}{l|l|l}
    \hline
    \bf{Method} & \textbf{Anchor} & \textbf{Linked Document (Manual  Snippet)} \\
    \hline
    \texttt{Discriminator} & oregon short line (+) & ...a rail line owned and operated by the union pacific railroad in the u.s. state of utah (-)... \\
    & new york state police (+) & ...2007 pictures published by the new york state police carefully (-)... \\
    & greenwich time (+) & ...search for : greenwich time greenwich sponsored links (-)... \\
    & pojoaque pueblo (+) & ...contact pojoaque pueblo here for your perusal is (+)... \\
    & banner ads (-) & ...online banner advertising blog - rupiz ads home about buzz (+)... \\
    \hline
    \texttt{\ris{}}
     & bmw 325i (+) & ...as soon as your order is finalized, your bmw 325i aftermarket clutch pivot pin is dispatched (+)... \\
     & bible bee (+) & ...general overview bible bee basics bee-attitudes statement of faith philosophy (+)... \\
     & kansas city zoo (+) & ...yahoo experience kansas city - kansas city's new zoo take a safari through africa (+)... \\
     & invasive alien species (+) & ...the north european and baltic network on invasive alien species (nobanis) is a gateway to (+)... \\
     & power load (-) & ...in the aliens movie, to combat the queen alien, ripley stepped into her a power loader (+)... \\
    \hline
    \end{tabular}
\end{table*}

%% file: Tables/case.tex
\begin{table}[h]
    \centering
    \caption{Examples of selected anchors and manually picked similar queries.
    \label{tab:case:anchor}}
    \small
    \begin{tabular}{l|l}
    \hline
    \bf{ClueWeb09-B Query} & \bf{Anchor} \\
    \hline
     dieting & crash dieting \\
     french lick resort and casino & tropicana casino $\&$ resort atlantic city \\
     diabetes education & vegan menu for people with diabetes \\
     income tax return online & personal income taxes \\
     orange county convention center & orange county convention center \\
    \hline
     \bf{Robust04 Query} & \bf{Anchor} \\
    \hline
     most dangerous vehicles & vehicle injury cases \\
     international art crime & art crimes \\
     mexican air pollution & outdoor air pollution \\
     commercial cyanide uses & cyanide pills \\
     el nino & el nino \\
    \hline
     \bf{ClueWeb12-B13 Query} & \bf{Anchor} \\
    \hline
     wind power & wind power in pa \\
     nba records & nba records \\
     teddy bears & beanie baby teddy bears \\
     benefits of yoga & benefits of yoga \\
     balding cure & balding treatment \\
    \hline
    \end{tabular}
\end{table}

%% file: conclusion.tex
\section{Conclusion}

\ris{} leverages the widely available anchor data to weakly supervise neural rankers and mitigates their dependency on large amounts of relevance labels.
To handle the noises in anchor data, \ris{} uses policy gradient to connect the demand---the needs of training signals from neural ranker---and the supply---anchor-document pairs, to select more effective weak supervision signals.
On three widely studied TREC benchmarks, \ris{} is the only weak supervision method that guides neural rankers stably outperform feature-based learning to rank methods. Using only publicly available data, it also nearly matches the effectiveness of training signals from private commercial search logs.

In our experiments, \ris{} robustly selects better anchor-document pairs than previous weak supervision approaches disconnected from target neural models. We also found that \ris{} tailors the weak supervision for each individual Neu-IR model as well as its convergence status. 
Intuitively, \ris{} starts with providing as much supervision as possible when the ranker is random; then it further elevates the neural ranker's performance using anchor-document pairs that well approximate query and relevant documents. 

\ris{} also provides a handy experiment ground to analyze the advantages and disadvantages of Neu-IR. We show results on how neural rankers work with feature-based learning to rank methods, what is the effective fine-tuning strategies on target ranking tasks, and how different training data amounts and query-document fractions influence Neu-IR models.
These analyses were hard to do as Neu-IR models were limited by the lack of large scale relevance-specific supervision.
\ris{} provides a simple way to lessen this dependency thus will facilitate and broaden the impact of deep learning research in information retrieval.

%% file: acknowledge.tex
\section{Acknowledgements}

Kaitao Zhang and Zhiyuan Liu are supported by the National Natural Science Foundation of China (NSFC61661146007, 61772302).
Si Sun, Houyu Zhang, Kaitao Zhang, and Zhenghao Liu conducted human evaluations.
We thank Guoqing Zheng, Paul Bennett, and Susan Dumais  for discussions in weak supervision methodologies,  the Anchor intuition, and the iterative view of pretrainig-application.
We thank Zhuyun Dai and Jamie Callan for sharing the base retrieval results on ClueWeb09-B and Robust04.

%% file: main.bbl

\begin{thebibliography}{54}


\ifx \showCODEN    \undefined \def \showCODEN     #1{\unskip}     \fi
\ifx \showDOI      \undefined \def \showDOI       #1{#1}\fi
\ifx \showISBNx    \undefined \def \showISBNx     #1{\unskip}     \fi
\ifx \showISBNxiii \undefined \def \showISBNxiii  #1{\unskip}     \fi
\ifx \showISSN     \undefined \def \showISSN      #1{\unskip}     \fi
\ifx \showLCCN     \undefined \def \showLCCN      #1{\unskip}     \fi
\ifx \shownote     \undefined \def \shownote      #1{#1}          \fi
\ifx \showarticletitle \undefined \def \showarticletitle #1{#1}   \fi
\ifx \showURL      \undefined \def \showURL       {\relax}        \fi
\providecommand\bibfield[2]{#2}
\providecommand\bibinfo[2]{#2}
\providecommand\natexlab[1]{#1}
\providecommand\showeprint[2][]{arXiv:#2}

\bibitem[\protect\citeauthoryear{Ahmad, Constant, Yang, and Cer}{Ahmad
  et~al\mbox{.}}{2019}]%
        {ahmad2019reqa}
\bibfield{author}{\bibinfo{person}{Amin Ahmad}, \bibinfo{person}{Noah
  Constant}, \bibinfo{person}{Yinfei Yang}, {and} \bibinfo{person}{Daniel
  Cer}.} \bibinfo{year}{2019}\natexlab{}.
\newblock \showarticletitle{ReQA: An Evaluation for End-to-End Answer Retrieval
  Models}.
\newblock \bibinfo{journal}{\emph{CoRR}}  \bibinfo{volume}{abs/1907.04780}
  (\bibinfo{year}{2019}).
\newblock


\bibitem[\protect\citeauthoryear{Berger and Lafferty}{Berger and
  Lafferty}{1999}]%
        {berger1999Information}
\bibfield{author}{\bibinfo{person}{Adam~L. Berger} {and}
  \bibinfo{person}{John~D. Lafferty}.} \bibinfo{year}{1999}\natexlab{}.
\newblock \showarticletitle{Information Retrieval as Statistical Translation}.
  In \bibinfo{booktitle}{\emph{{SIGIR} '99: Proceedings of the 22nd Annual
  International {ACM} {SIGIR} Conference on Research and Development in
  Information Retrieval, August 15-19, 1999, Berkeley, CA, {USA}}}.
  \bibinfo{pages}{222--229}.
\newblock


\bibitem[\protect\citeauthoryear{Croft, Metzler, and Strohman}{Croft
  et~al\mbox{.}}{2009}]%
        {croft2010search}
\bibfield{author}{\bibinfo{person}{W.~Bruce Croft}, \bibinfo{person}{Donald
  Metzler}, {and} \bibinfo{person}{Trevor Strohman}.}
  \bibinfo{year}{2009}\natexlab{}.
\newblock \bibinfo{booktitle}{\emph{Search Engines - Information Retrieval in
  Practice}}.
\newblock \bibinfo{publisher}{Pearson Education}.
\newblock


\bibitem[\protect\citeauthoryear{Dai and Callan}{Dai and Callan}{2019}]%
        {dai2019deeper}
\bibfield{author}{\bibinfo{person}{Zhuyun Dai} {and} \bibinfo{person}{Jamie
  Callan}.} \bibinfo{year}{2019}\natexlab{}.
\newblock \showarticletitle{Deeper Text Understanding for {IR} with Contextual
  Neural Language Modeling}. In \bibinfo{booktitle}{\emph{Proceedings of the
  42nd International {ACM} {SIGIR} Conference on Research and Development in
  Information Retrieval, {SIGIR} 2019, Paris, France, July 21-25, 2019}}.
  \bibinfo{pages}{985--988}.
\newblock


\bibitem[\protect\citeauthoryear{Dai, Xiong, Callan, and Liu}{Dai
  et~al\mbox{.}}{2018}]%
        {convknrm}
\bibfield{author}{\bibinfo{person}{Zhuyun Dai}, \bibinfo{person}{Chenyan
  Xiong}, \bibinfo{person}{Jamie Callan}, {and} \bibinfo{person}{Zhiyuan Liu}.}
  \bibinfo{year}{2018}\natexlab{}.
\newblock \showarticletitle{Convolutional Neural Networks for Soft-Matching
  N-Grams in Ad-hoc Search}. In \bibinfo{booktitle}{\emph{Proceedings of the
  Eleventh {ACM} International Conference on Web Search and Data Mining, {WSDM}
  2018, Marina Del Rey, CA, USA, February 5-9, 2018}}.
  \bibinfo{pages}{126--134}.
\newblock


\bibitem[\protect\citeauthoryear{Dang and Croft}{Dang and Croft}{2010}]%
        {dang2010query}
\bibfield{author}{\bibinfo{person}{Van Dang} {and} \bibinfo{person}{W.~Bruce
  Croft}.} \bibinfo{year}{2010}\natexlab{}.
\newblock \showarticletitle{Query reformulation using anchor text}. In
  \bibinfo{booktitle}{\emph{Proceedings of the Third International Conference
  on Web Search and Web Data Mining, {WSDM} 2010, New York, NY, USA, February
  4-6, 2010}}. \bibinfo{pages}{41--50}.
\newblock


\bibitem[\protect\citeauthoryear{Dehghani, Severyn, Rothe, and Kamps}{Dehghani
  et~al\mbox{.}}{2017a}]%
        {dehghani2017learning}
\bibfield{author}{\bibinfo{person}{Mostafa Dehghani}, \bibinfo{person}{Aliaksei
  Severyn}, \bibinfo{person}{Sascha Rothe}, {and} \bibinfo{person}{Jaap
  Kamps}.} \bibinfo{year}{2017}\natexlab{a}.
\newblock \showarticletitle{Learning to Learn from Weak Supervision by Full
  Supervision}.
\newblock \bibinfo{journal}{\emph{CoRR}}  \bibinfo{volume}{abs/1711.11383}
  (\bibinfo{year}{2017}).
\newblock


\bibitem[\protect\citeauthoryear{Dehghani, Zamani, Severyn, Kamps, and
  Croft}{Dehghani et~al\mbox{.}}{2017b}]%
        {dehghani2017neural}
\bibfield{author}{\bibinfo{person}{Mostafa Dehghani}, \bibinfo{person}{Hamed
  Zamani}, \bibinfo{person}{Aliaksei Severyn}, \bibinfo{person}{Jaap Kamps},
  {and} \bibinfo{person}{W.~Bruce Croft}.} \bibinfo{year}{2017}\natexlab{b}.
\newblock \showarticletitle{Neural Ranking Models with Weak Supervision}. In
  \bibinfo{booktitle}{\emph{Proceedings of the 40th International {ACM} {SIGIR}
  Conference on Research and Development in Information Retrieval, Shinjuku,
  Tokyo, Japan, August 7-11, 2017}}. \bibinfo{pages}{65--74}.
\newblock


\bibitem[\protect\citeauthoryear{Devlin, Chang, Lee, and Toutanova}{Devlin
  et~al\mbox{.}}{2019}]%
        {devlin2019bert}
\bibfield{author}{\bibinfo{person}{Jacob Devlin}, \bibinfo{person}{Ming{-}Wei
  Chang}, \bibinfo{person}{Kenton Lee}, {and} \bibinfo{person}{Kristina
  Toutanova}.} \bibinfo{year}{2019}\natexlab{}.
\newblock \showarticletitle{{BERT:} Pre-training of Deep Bidirectional
  Transformers for Language Understanding}. In
  \bibinfo{booktitle}{\emph{Proceedings of the 2019 Conference of the North
  American Chapter of the Association for Computational Linguistics: Human
  Language Technologies, {NAACL-HLT} 2019, Minneapolis, MN, USA, June 2-7,
  2019, Volume 1 (Long and Short Papers)}}. \bibinfo{pages}{4171--4186}.
\newblock


\bibitem[\protect\citeauthoryear{Diaz, Mitra, and Craswell}{Diaz
  et~al\mbox{.}}{2016}]%
        {diaz2016query}
\bibfield{author}{\bibinfo{person}{Fernando Diaz}, \bibinfo{person}{Bhaskar
  Mitra}, {and} \bibinfo{person}{Nick Craswell}.}
  \bibinfo{year}{2016}\natexlab{}.
\newblock \showarticletitle{Query Expansion with Locally-Trained Word
  Embeddings}. In \bibinfo{booktitle}{\emph{Proceedings of the 54th Annual
  Meeting of the Association for Computational Linguistics, {ACL} 2016, August
  7-12, 2016, Berlin, Germany, Volume 1: Long Papers}}.
\newblock


\bibitem[\protect\citeauthoryear{Dou, Song, Nie, and Wen}{Dou
  et~al\mbox{.}}{2009}]%
        {dou2009using}
\bibfield{author}{\bibinfo{person}{Zhicheng Dou}, \bibinfo{person}{Ruihua
  Song}, \bibinfo{person}{Jian{-}Yun Nie}, {and} \bibinfo{person}{Ji{-}Rong
  Wen}.} \bibinfo{year}{2009}\natexlab{}.
\newblock \showarticletitle{Using anchor texts with their hyperlink structure
  for web search}. In \bibinfo{booktitle}{\emph{Proceedings of the 32nd Annual
  International {ACM} {SIGIR} Conference on Research and Development in
  Information Retrieval, {SIGIR} 2009, Boston, MA, USA, July 19-23, 2009}}.
  \bibinfo{pages}{227--234}.
\newblock


\bibitem[\protect\citeauthoryear{Eiron and McCurley}{Eiron and
  McCurley}{2003}]%
        {eiron2003analysis}
\bibfield{author}{\bibinfo{person}{Nadav Eiron} {and} \bibinfo{person}{Kevin~S.
  McCurley}.} \bibinfo{year}{2003}\natexlab{}.
\newblock \showarticletitle{Analysis of anchor text for web search}. In
  \bibinfo{booktitle}{\emph{{SIGIR} 2003: Proceedings of the 26th Annual
  International {ACM} {SIGIR} Conference on Research and Development in
  Information Retrieval, July 28 - August 1, 2003, Toronto, Canada}}.
  \bibinfo{pages}{459--460}.
\newblock


\bibitem[\protect\citeauthoryear{Guo, Fan, Ai, and Croft}{Guo
  et~al\mbox{.}}{2016}]%
        {jiafeng2016deep}
\bibfield{author}{\bibinfo{person}{Jiafeng Guo}, \bibinfo{person}{Yixing Fan},
  \bibinfo{person}{Qingyao Ai}, {and} \bibinfo{person}{W.~Bruce Croft}.}
  \bibinfo{year}{2016}\natexlab{}.
\newblock \showarticletitle{A Deep Relevance Matching Model for Ad-hoc
  Retrieval}. In \bibinfo{booktitle}{\emph{Proceedings of the 25th {ACM}
  International Conference on Information and Knowledge Management, {CIKM}
  2016, Indianapolis, IN, USA, October 24-28, 2016}}. \bibinfo{pages}{55--64}.
\newblock


\bibitem[\protect\citeauthoryear{Guo, Fan, Pang, Yang, Ai, Zamani, Wu, Croft,
  and Cheng}{Guo et~al\mbox{.}}{2019}]%
        {guo2019deep}
\bibfield{author}{\bibinfo{person}{Jiafeng Guo}, \bibinfo{person}{Yixing Fan},
  \bibinfo{person}{Liang Pang}, \bibinfo{person}{Liu Yang},
  \bibinfo{person}{Qingyao Ai}, \bibinfo{person}{Hamed Zamani},
  \bibinfo{person}{Chen Wu}, \bibinfo{person}{W.~Bruce Croft}, {and}
  \bibinfo{person}{Xueqi Cheng}.} \bibinfo{year}{2019}\natexlab{}.
\newblock \showarticletitle{A Deep Look into Neural Ranking Models for
  Information Retrieval}.
\newblock \bibinfo{journal}{\emph{CoRR}}  \bibinfo{volume}{abs/1903.06902}
  (\bibinfo{year}{2019}).
\newblock


\bibitem[\protect\citeauthoryear{Han, Yao, Yu, Niu, Xu, Hu, Tsang, and
  Sugiyama}{Han et~al\mbox{.}}{2018}]%
        {han2018co}
\bibfield{author}{\bibinfo{person}{Bo Han}, \bibinfo{person}{Quanming Yao},
  \bibinfo{person}{Xingrui Yu}, \bibinfo{person}{Gang Niu},
  \bibinfo{person}{Miao Xu}, \bibinfo{person}{Weihua Hu},
  \bibinfo{person}{Ivor~W. Tsang}, {and} \bibinfo{person}{Masashi Sugiyama}.}
  \bibinfo{year}{2018}\natexlab{}.
\newblock \showarticletitle{Co-teaching: Robust training of deep neural
  networks with extremely noisy labels}. In \bibinfo{booktitle}{\emph{Advances
  in Neural Information Processing Systems 31: Annual Conference on Neural
  Information Processing Systems 2018, NeurIPS 2018, 3-8 December 2018,
  Montr{\'{e}}al, Canada}}. \bibinfo{pages}{8536--8546}.
\newblock


\bibitem[\protect\citeauthoryear{Hendrycks, Mazeika, Wilson, and
  Gimpel}{Hendrycks et~al\mbox{.}}{2018}]%
        {hendrycks2018using}
\bibfield{author}{\bibinfo{person}{Dan Hendrycks}, \bibinfo{person}{Mantas
  Mazeika}, \bibinfo{person}{Duncan Wilson}, {and} \bibinfo{person}{Kevin
  Gimpel}.} \bibinfo{year}{2018}\natexlab{}.
\newblock \showarticletitle{Using Trusted Data to Train Deep Networks on Labels
  Corrupted by Severe Noise}. In \bibinfo{booktitle}{\emph{Advances in Neural
  Information Processing Systems 31: Annual Conference on Neural Information
  Processing Systems 2018, NeurIPS 2018, 3-8 December 2018, Montr{\'{e}}al,
  Canada}}. \bibinfo{pages}{10477--10486}.
\newblock


\bibitem[\protect\citeauthoryear{Hofst{\"{a}}tter, Rekabsaz, Eickhoff, and
  Hanbury}{Hofst{\"{a}}tter et~al\mbox{.}}{2019}]%
        {hofstatter2019effect}
\bibfield{author}{\bibinfo{person}{Sebastian Hofst{\"{a}}tter},
  \bibinfo{person}{Navid Rekabsaz}, \bibinfo{person}{Carsten Eickhoff}, {and}
  \bibinfo{person}{Allan Hanbury}.} \bibinfo{year}{2019}\natexlab{}.
\newblock \showarticletitle{On the Effect of Low-Frequency Terms on Neural-IR
  Models}. In \bibinfo{booktitle}{\emph{Proceedings of the 42nd International
  {ACM} {SIGIR} Conference on Research and Development in Information
  Retrieval, {SIGIR} 2019, Paris, France, July 21-25, 2019}}.
  \bibinfo{pages}{1137--1140}.
\newblock


\bibitem[\protect\citeauthoryear{Hu, Lu, Li, and Chen}{Hu
  et~al\mbox{.}}{2014}]%
        {arcii}
\bibfield{author}{\bibinfo{person}{Baotian Hu}, \bibinfo{person}{Zhengdong Lu},
  \bibinfo{person}{Hang Li}, {and} \bibinfo{person}{Qingcai Chen}.}
  \bibinfo{year}{2014}\natexlab{}.
\newblock \showarticletitle{Convolutional Neural Network Architectures for
  Matching Natural Language Sentences}. In \bibinfo{booktitle}{\emph{Advances
  in Neural Information Processing Systems 27: Annual Conference on Neural
  Information Processing Systems 2014, December 8-13 2014, Montreal, Quebec,
  Canada}}. \bibinfo{pages}{2042--2050}.
\newblock


\bibitem[\protect\citeauthoryear{Huang, He, Gao, Deng, Acero, and Heck}{Huang
  et~al\mbox{.}}{2013}]%
        {huang2013learning}
\bibfield{author}{\bibinfo{person}{Po{-}Sen Huang}, \bibinfo{person}{Xiaodong
  He}, \bibinfo{person}{Jianfeng Gao}, \bibinfo{person}{Li Deng},
  \bibinfo{person}{Alex Acero}, {and} \bibinfo{person}{Larry~P. Heck}.}
  \bibinfo{year}{2013}\natexlab{}.
\newblock \showarticletitle{Learning deep structured semantic models for web
  search using clickthrough data}. In \bibinfo{booktitle}{\emph{22nd {ACM}
  International Conference on Information and Knowledge Management, CIKM'13,
  San Francisco, CA, USA, October 27 - November 1, 2013}}.
  \bibinfo{pages}{2333--2338}.
\newblock


\bibitem[\protect\citeauthoryear{Hui, Yates, Berberich, and de~Melo}{Hui
  et~al\mbox{.}}{2017}]%
        {hui2017pacrr}
\bibfield{author}{\bibinfo{person}{Kai Hui}, \bibinfo{person}{Andrew Yates},
  \bibinfo{person}{Klaus Berberich}, {and} \bibinfo{person}{Gerard de Melo}.}
  \bibinfo{year}{2017}\natexlab{}.
\newblock \showarticletitle{{PACRR:} {A} Position-Aware Neural {IR} Model for
  Relevance Matching}. In \bibinfo{booktitle}{\emph{Proceedings of the 2017
  Conference on Empirical Methods in Natural Language Processing, {EMNLP} 2017,
  Copenhagen, Denmark, September 9-11, 2017}}. \bibinfo{pages}{1049--1058}.
\newblock


\bibitem[\protect\citeauthoryear{Kraft and Zien}{Kraft and Zien}{2004}]%
        {kraft2004mining}
\bibfield{author}{\bibinfo{person}{Reiner Kraft} {and}
  \bibinfo{person}{Jason~Y. Zien}.} \bibinfo{year}{2004}\natexlab{}.
\newblock \showarticletitle{Mining anchor text for query refinement}. In
  \bibinfo{booktitle}{\emph{Proceedings of the 13th international conference on
  World Wide Web, {WWW} 2004, New York, NY, USA, May 17-20, 2004}}.
  \bibinfo{pages}{666--674}.
\newblock


\bibitem[\protect\citeauthoryear{Lee, Chang, and Toutanova}{Lee
  et~al\mbox{.}}{2019}]%
        {lee2019latent}
\bibfield{author}{\bibinfo{person}{Kenton Lee}, \bibinfo{person}{Ming{-}Wei
  Chang}, {and} \bibinfo{person}{Kristina Toutanova}.}
  \bibinfo{year}{2019}\natexlab{}.
\newblock \showarticletitle{Latent Retrieval for Weakly Supervised Open Domain
  Question Answering}. In \bibinfo{booktitle}{\emph{Proceedings of the 57th
  Conference of the Association for Computational Linguistics, {ACL} 2019,
  Florence, Italy, July 28- August 2, 2019, Volume 1: Long Papers}}.
  \bibinfo{pages}{6086--6096}.
\newblock


\bibitem[\protect\citeauthoryear{Liu, Xiong, Sun, and Liu}{Liu
  et~al\mbox{.}}{2018}]%
        {liu2018entity}
\bibfield{author}{\bibinfo{person}{Zheng{-}Hao Liu}, \bibinfo{person}{Chenyan
  Xiong}, \bibinfo{person}{Maosong Sun}, {and} \bibinfo{person}{Zhiyuan Liu}.}
  \bibinfo{year}{2018}\natexlab{}.
\newblock \showarticletitle{Entity-Duet Neural Ranking: Understanding the Role
  of Knowledge Graph Semantics in Neural Information Retrieval}. In
  \bibinfo{booktitle}{\emph{Proceedings of the 56th Annual Meeting of the
  Association for Computational Linguistics, {ACL} 2018, Melbourne, Australia,
  July 15-20, 2018, Volume 1: Long Papers}}. \bibinfo{pages}{2395--2405}.
\newblock


\bibitem[\protect\citeauthoryear{Luo, Zheng, Mao, Liu, Zhang, and Ma}{Luo
  et~al\mbox{.}}{2017}]%
        {luo2017training}
\bibfield{author}{\bibinfo{person}{Cheng Luo}, \bibinfo{person}{Yukun Zheng},
  \bibinfo{person}{Jiaxin Mao}, \bibinfo{person}{Yiqun Liu},
  \bibinfo{person}{Min Zhang}, {and} \bibinfo{person}{Shaoping Ma}.}
  \bibinfo{year}{2017}\natexlab{}.
\newblock \showarticletitle{Training Deep Ranking Model with Weak Relevance
  Labels}. In \bibinfo{booktitle}{\emph{Databases Theory and Applications -
  28th Australasian Database Conference, {ADC} 2017, Brisbane, QLD, Australia,
  September 25-28, 2017, Proceedings}}. \bibinfo{pages}{205--216}.
\newblock


\bibitem[\protect\citeauthoryear{MacAvaney, Yates, Cohan, and
  Goharian}{MacAvaney et~al\mbox{.}}{2019a}]%
        {macavaney2019cedr}
\bibfield{author}{\bibinfo{person}{Sean MacAvaney}, \bibinfo{person}{Andrew
  Yates}, \bibinfo{person}{Arman Cohan}, {and} \bibinfo{person}{Nazli
  Goharian}.} \bibinfo{year}{2019}\natexlab{a}.
\newblock \showarticletitle{{CEDR:} Contextualized Embeddings for Document
  Ranking}. In \bibinfo{booktitle}{\emph{Proceedings of the 42nd International
  {ACM} {SIGIR} Conference on Research and Development in Information
  Retrieval, {SIGIR} 2019, Paris, France, July 21-25, 2019}}.
  \bibinfo{pages}{1101--1104}.
\newblock


\bibitem[\protect\citeauthoryear{MacAvaney, Yates, Hui, and Frieder}{MacAvaney
  et~al\mbox{.}}{2019b}]%
        {macavaney2019content}
\bibfield{author}{\bibinfo{person}{Sean MacAvaney}, \bibinfo{person}{Andrew
  Yates}, \bibinfo{person}{Kai Hui}, {and} \bibinfo{person}{Ophir Frieder}.}
  \bibinfo{year}{2019}\natexlab{b}.
\newblock \showarticletitle{Content-Based Weak Supervision for Ad-Hoc
  Re-Ranking}. In \bibinfo{booktitle}{\emph{Proceedings of the 42nd
  International {ACM} {SIGIR} Conference on Research and Development in
  Information Retrieval, {SIGIR} 2019, Paris, France, July 21-25, 2019}}.
  \bibinfo{pages}{993--996}.
\newblock


\bibitem[\protect\citeauthoryear{Mikolov, Sutskever, Chen, Corrado, and
  Dean}{Mikolov et~al\mbox{.}}{2013}]%
        {word2vec}
\bibfield{author}{\bibinfo{person}{Tomas Mikolov}, \bibinfo{person}{Ilya
  Sutskever}, \bibinfo{person}{Kai Chen}, \bibinfo{person}{Gregory~S. Corrado},
  {and} \bibinfo{person}{Jeffrey Dean}.} \bibinfo{year}{2013}\natexlab{}.
\newblock \showarticletitle{Distributed Representations of Words and Phrases
  and their Compositionality}. In \bibinfo{booktitle}{\emph{Advances in Neural
  Information Processing Systems 26: 27th Annual Conference on Neural
  Information Processing Systems 2013. Proceedings of a meeting held December
  5-8, 2013, Lake Tahoe, Nevada, United States}}. \bibinfo{pages}{3111--3119}.
\newblock


\bibitem[\protect\citeauthoryear{Nguyen, Rosenberg, Song, Gao, Tiwary,
  Majumder, and Deng}{Nguyen et~al\mbox{.}}{2016}]%
        {msmarco}
\bibfield{author}{\bibinfo{person}{Tri Nguyen}, \bibinfo{person}{Mir
  Rosenberg}, \bibinfo{person}{Xia Song}, \bibinfo{person}{Jianfeng Gao},
  \bibinfo{person}{Saurabh Tiwary}, \bibinfo{person}{Rangan Majumder}, {and}
  \bibinfo{person}{Li Deng}.} \bibinfo{year}{2016}\natexlab{}.
\newblock \showarticletitle{{MS} {MARCO:} {A} Human Generated MAchine Reading
  COmprehension Dataset}. In \bibinfo{booktitle}{\emph{Proceedings of the
  Workshop on Cognitive Computation: Integrating neural and symbolic approaches
  2016 co-located with the 30th Annual Conference on Neural Information
  Processing Systems {(NIPS} 2016), Barcelona, Spain, December 9, 2016}}.
\newblock


\bibitem[\protect\citeauthoryear{Nogueira and Cho}{Nogueira and Cho}{2019}]%
        {nogueira2019passage}
\bibfield{author}{\bibinfo{person}{Rodrigo Nogueira} {and}
  \bibinfo{person}{Kyunghyun Cho}.} \bibinfo{year}{2019}\natexlab{}.
\newblock \showarticletitle{Passage Re-ranking with {BERT}}.
\newblock \bibinfo{journal}{\emph{CoRR}}  \bibinfo{volume}{abs/1901.04085}
  (\bibinfo{year}{2019}).
\newblock


\bibitem[\protect\citeauthoryear{Nogueira, Yang, Lin, and Cho}{Nogueira
  et~al\mbox{.}}{2019}]%
        {nogueira2019document}
\bibfield{author}{\bibinfo{person}{Rodrigo Nogueira}, \bibinfo{person}{Wei
  Yang}, \bibinfo{person}{Jimmy Lin}, {and} \bibinfo{person}{Kyunghyun Cho}.}
  \bibinfo{year}{2019}\natexlab{}.
\newblock \showarticletitle{Document Expansion by Query Prediction}.
\newblock \bibinfo{journal}{\emph{CoRR}}  \bibinfo{volume}{abs/1904.08375}
  (\bibinfo{year}{2019}).
\newblock


\bibitem[\protect\citeauthoryear{Padigela, Zamani, and Croft}{Padigela
  et~al\mbox{.}}{2019}]%
        {padigela2019investigating}
\bibfield{author}{\bibinfo{person}{Harshith Padigela}, \bibinfo{person}{Hamed
  Zamani}, {and} \bibinfo{person}{W.~Bruce Croft}.}
  \bibinfo{year}{2019}\natexlab{}.
\newblock \showarticletitle{Investigating the Successes and Failures of {BERT}
  for Passage Re-Ranking}.
\newblock \bibinfo{journal}{\emph{CoRR}}  \bibinfo{volume}{abs/1905.01758}
  (\bibinfo{year}{2019}).
\newblock


\bibitem[\protect\citeauthoryear{Pang, Lan, Guo, Xu, Wan, and Cheng}{Pang
  et~al\mbox{.}}{2016}]%
        {Pang2016TextMA}
\bibfield{author}{\bibinfo{person}{Liang Pang}, \bibinfo{person}{Yanyan Lan},
  \bibinfo{person}{Jiafeng Guo}, \bibinfo{person}{Jun Xu},
  \bibinfo{person}{Shengxian Wan}, {and} \bibinfo{person}{Xueqi Cheng}.}
  \bibinfo{year}{2016}\natexlab{}.
\newblock \bibinfo{title}{Text Matching as Image Recognition}.
\newblock , \bibinfo{numpages}{2793--2799}~pages.
\newblock


\bibitem[\protect\citeauthoryear{Pang, Lan, Guo, Xu, Xu, and Cheng}{Pang
  et~al\mbox{.}}{2017}]%
        {pang2017deeprank}
\bibfield{author}{\bibinfo{person}{Liang Pang}, \bibinfo{person}{Yanyan Lan},
  \bibinfo{person}{Jiafeng Guo}, \bibinfo{person}{Jun Xu},
  \bibinfo{person}{Jingfang Xu}, {and} \bibinfo{person}{Xueqi Cheng}.}
  \bibinfo{year}{2017}\natexlab{}.
\newblock \showarticletitle{DeepRank: {A} New Deep Architecture for Relevance
  Ranking in Information Retrieval}. In \bibinfo{booktitle}{\emph{Proceedings
  of the 2017 {ACM} on Conference on Information and Knowledge Management,
  {CIKM} 2017, Singapore, November 06 - 10, 2017}}. \bibinfo{pages}{257--266}.
\newblock


\bibitem[\protect\citeauthoryear{Pennington, Socher, and Manning}{Pennington
  et~al\mbox{.}}{2014}]%
        {pennington2014glove}
\bibfield{author}{\bibinfo{person}{Jeffrey Pennington},
  \bibinfo{person}{Richard Socher}, {and} \bibinfo{person}{Christopher~D.
  Manning}.} \bibinfo{year}{2014}\natexlab{}.
\newblock \showarticletitle{Glove: Global Vectors for Word Representation}. In
  \bibinfo{booktitle}{\emph{Proceedings of the 2014 Conference on Empirical
  Methods in Natural Language Processing, {EMNLP} 2014, October 25-29, 2014,
  Doha, Qatar, {A} meeting of SIGDAT, a Special Interest Group of the {ACL}}}.
  \bibinfo{pages}{1532--1543}.
\newblock


\bibitem[\protect\citeauthoryear{Peters, Neumann, Iyyer, Gardner, Clark, Lee,
  and Zettlemoyer}{Peters et~al\mbox{.}}{2018}]%
        {PetersELMO}
\bibfield{author}{\bibinfo{person}{Matthew~E. Peters}, \bibinfo{person}{Mark
  Neumann}, \bibinfo{person}{Mohit Iyyer}, \bibinfo{person}{Matt Gardner},
  \bibinfo{person}{Christopher Clark}, \bibinfo{person}{Kenton Lee}, {and}
  \bibinfo{person}{Luke Zettlemoyer}.} \bibinfo{year}{2018}\natexlab{}.
\newblock \showarticletitle{Deep Contextualized Word Representations}. In
  \bibinfo{booktitle}{\emph{Proceedings of the 2018 Conference of the North
  American Chapter of the Association for Computational Linguistics: Human
  Language Technologies, {NAACL-HLT} 2018, New Orleans, Louisiana, USA, June
  1-6, 2018, Volume 1 (Long Papers)}}. \bibinfo{pages}{2227--2237}.
\newblock


\bibitem[\protect\citeauthoryear{Pyreddy, Ramaseshan, Joshi, Dai, Xiong,
  Callan, and Liu}{Pyreddy et~al\mbox{.}}{2018}]%
        {pyreddy2018consistency}
\bibfield{author}{\bibinfo{person}{Mary~Arpita Pyreddy},
  \bibinfo{person}{Varshini Ramaseshan}, \bibinfo{person}{Narendra~Nath Joshi},
  \bibinfo{person}{Zhuyun Dai}, \bibinfo{person}{Chenyan Xiong},
  \bibinfo{person}{Jamie Callan}, {and} \bibinfo{person}{Zhiyuan Liu}.}
  \bibinfo{year}{2018}\natexlab{}.
\newblock \showarticletitle{Consistency and Variation in Kernel Neural Ranking
  Model}. In \bibinfo{booktitle}{\emph{The 41st International {ACM} {SIGIR}
  Conference on Research {\&} Development in Information Retrieval, {SIGIR}
  2018, Ann Arbor, MI, USA, July 08-12, 2018}}. \bibinfo{pages}{961--964}.
\newblock


\bibitem[\protect\citeauthoryear{Qiao, Xiong, Liu, and Liu}{Qiao
  et~al\mbox{.}}{2019}]%
        {qiao2019understanding}
\bibfield{author}{\bibinfo{person}{Yifan Qiao}, \bibinfo{person}{Chenyan
  Xiong}, \bibinfo{person}{Zheng{-}Hao Liu}, {and} \bibinfo{person}{Zhiyuan
  Liu}.} \bibinfo{year}{2019}\natexlab{}.
\newblock \showarticletitle{Understanding the Behaviors of {BERT} in Ranking}.
\newblock \bibinfo{journal}{\emph{CoRR}}  \bibinfo{volume}{abs/1904.07531}
  (\bibinfo{year}{2019}).
\newblock


\bibitem[\protect\citeauthoryear{Qu, Ji, Qiu, Yang, Min, Chen, Huang, and
  Croft}{Qu et~al\mbox{.}}{2019}]%
        {qu2019learning}
\bibfield{author}{\bibinfo{person}{Chen Qu}, \bibinfo{person}{Feng Ji},
  \bibinfo{person}{Minghui Qiu}, \bibinfo{person}{Liu Yang},
  \bibinfo{person}{Zhiyu Min}, \bibinfo{person}{Haiqing Chen},
  \bibinfo{person}{Jun Huang}, {and} \bibinfo{person}{W.~Bruce Croft}.}
  \bibinfo{year}{2019}\natexlab{}.
\newblock \showarticletitle{Learning to Selectively Transfer: Reinforced
  Transfer Learning for Deep Text Matching}. In
  \bibinfo{booktitle}{\emph{Proceedings of the Twelfth {ACM} International
  Conference on Web Search and Data Mining, {WSDM} 2019, Melbourne, VIC,
  Australia, February 11-15, 2019}}. \bibinfo{pages}{699--707}.
\newblock


\bibitem[\protect\citeauthoryear{Ren, Zeng, Yang, and Urtasun}{Ren
  et~al\mbox{.}}{2018}]%
        {ren2018learning}
\bibfield{author}{\bibinfo{person}{Mengye Ren}, \bibinfo{person}{Wenyuan Zeng},
  \bibinfo{person}{Bin Yang}, {and} \bibinfo{person}{Raquel Urtasun}.}
  \bibinfo{year}{2018}\natexlab{}.
\newblock \showarticletitle{Learning to Reweight Examples for Robust Deep
  Learning}. In \bibinfo{booktitle}{\emph{Proceedings of the 35th International
  Conference on Machine Learning, {ICML} 2018, Stockholmsm{\"{a}}ssan,
  Stockholm, Sweden, July 10-15, 2018}}. \bibinfo{pages}{4331--4340}.
\newblock


\bibitem[\protect\citeauthoryear{Shen, He, Gao, Deng, and Mesnil}{Shen
  et~al\mbox{.}}{2014}]%
        {cdssm}
\bibfield{author}{\bibinfo{person}{Yelong Shen}, \bibinfo{person}{Xiaodong He},
  \bibinfo{person}{Jianfeng Gao}, \bibinfo{person}{Li Deng}, {and}
  \bibinfo{person}{Gr{\'{e}}goire Mesnil}.} \bibinfo{year}{2014}\natexlab{}.
\newblock \showarticletitle{A Latent Semantic Model with Convolutional-Pooling
  Structure for Information Retrieval}. In
  \bibinfo{booktitle}{\emph{Proceedings of the 23rd {ACM} International
  Conference on Conference on Information and Knowledge Management, {CIKM}
  2014, Shanghai, China, November 3-7, 2014}}. \bibinfo{pages}{101--110}.
\newblock


\bibitem[\protect\citeauthoryear{Wang, Qiu, Wang, Li, Gong, Zeng, Huang, Zheng,
  Cai, and Zhou}{Wang et~al\mbox{.}}{2019}]%
        {wang2019minimax}
\bibfield{author}{\bibinfo{person}{Bo Wang}, \bibinfo{person}{Minghui Qiu},
  \bibinfo{person}{Xisen Wang}, \bibinfo{person}{Yaliang Li},
  \bibinfo{person}{Yu Gong}, \bibinfo{person}{Xiaoyi Zeng},
  \bibinfo{person}{Jun Huang}, \bibinfo{person}{Bo Zheng},
  \bibinfo{person}{Deng Cai}, {and} \bibinfo{person}{Jingren Zhou}.}
  \bibinfo{year}{2019}\natexlab{}.
\newblock \showarticletitle{A Minimax Game for Instance based Selective
  Transfer Learning}. In \bibinfo{booktitle}{\emph{Proceedings of the 25th
  {ACM} {SIGKDD} International Conference on Knowledge Discovery {\&} Data
  Mining, {KDD} 2019, Anchorage, AK, USA, August 4-8, 2019}}.
  \bibinfo{pages}{34--43}.
\newblock


\bibitem[\protect\citeauthoryear{Williams}{Williams}{1992}]%
        {williams1992simple}
\bibfield{author}{\bibinfo{person}{Ronald~J. Williams}.}
  \bibinfo{year}{1992}\natexlab{}.
\newblock \showarticletitle{Simple Statistical Gradient-Following Algorithms
  for Connectionist Reinforcement Learning}.
\newblock \bibinfo{journal}{\emph{Machine Learning}}  \bibinfo{volume}{8}
  (\bibinfo{year}{1992}), \bibinfo{pages}{229--256}.
\newblock


\bibitem[\protect\citeauthoryear{Xia, Xu, Lan, Guo, Zeng, and Cheng}{Xia
  et~al\mbox{.}}{2017}]%
        {xia2017adapting}
\bibfield{author}{\bibinfo{person}{Long Xia}, \bibinfo{person}{Jun Xu},
  \bibinfo{person}{Yanyan Lan}, \bibinfo{person}{Jiafeng Guo},
  \bibinfo{person}{Wei Zeng}, {and} \bibinfo{person}{Xueqi Cheng}.}
  \bibinfo{year}{2017}\natexlab{}.
\newblock \showarticletitle{Adapting Markov Decision Process for Search Result
  Diversification}. In \bibinfo{booktitle}{\emph{Proceedings of the 40th
  International {ACM} {SIGIR} Conference on Research and Development in
  Information Retrieval, Shinjuku, Tokyo, Japan, August 7-11, 2017}}.
  \bibinfo{pages}{535--544}.
\newblock


\bibitem[\protect\citeauthoryear{Xiong, Callan, and Liu}{Xiong
  et~al\mbox{.}}{2017a}]%
        {xiong2017duet}
\bibfield{author}{\bibinfo{person}{Chenyan Xiong}, \bibinfo{person}{Jamie
  Callan}, {and} \bibinfo{person}{Tie{-}Yan Liu}.}
  \bibinfo{year}{2017}\natexlab{a}.
\newblock \showarticletitle{Word-Entity Duet Representations for Document
  Ranking}. In \bibinfo{booktitle}{\emph{Proceedings of the 40th International
  {ACM} {SIGIR} Conference on Research and Development in Information
  Retrieval, Shinjuku, Tokyo, Japan, August 7-11, 2017}}.
  \bibinfo{pages}{763--772}.
\newblock


\bibitem[\protect\citeauthoryear{Xiong, Dai, Callan, Liu, and Power}{Xiong
  et~al\mbox{.}}{2017b}]%
        {xiong2017knrm}
\bibfield{author}{\bibinfo{person}{Chenyan Xiong}, \bibinfo{person}{Zhuyun
  Dai}, \bibinfo{person}{Jamie Callan}, \bibinfo{person}{Zhiyuan Liu}, {and}
  \bibinfo{person}{Russell Power}.} \bibinfo{year}{2017}\natexlab{b}.
\newblock \showarticletitle{End-to-End Neural Ad-hoc Ranking with Kernel
  Pooling}. In \bibinfo{booktitle}{\emph{Proceedings of the 40th International
  {ACM} {SIGIR} Conference on Research and Development in Information
  Retrieval, Shinjuku, Tokyo, Japan, August 7-11, 2017}}.
  \bibinfo{pages}{55--64}.
\newblock


\bibitem[\protect\citeauthoryear{Xiong, Liu, Callan, and Liu}{Xiong
  et~al\mbox{.}}{2018}]%
        {xiong2018towards}
\bibfield{author}{\bibinfo{person}{Chenyan Xiong}, \bibinfo{person}{Zhengzhong
  Liu}, \bibinfo{person}{Jamie Callan}, {and} \bibinfo{person}{Tie{-}Yan Liu}.}
  \bibinfo{year}{2018}\natexlab{}.
\newblock \showarticletitle{Towards Better Text Understanding and Retrieval
  through Kernel Entity Salience Modeling}. In \bibinfo{booktitle}{\emph{The
  41st International {ACM} {SIGIR} Conference on Research {\&} Development in
  Information Retrieval, {SIGIR} 2018, Ann Arbor, MI, USA, July 08-12, 2018}}.
  \bibinfo{pages}{575--584}.
\newblock


\bibitem[\protect\citeauthoryear{Yang, Lu, Yang, and Lin}{Yang
  et~al\mbox{.}}{2019}]%
        {yang2019critically}
\bibfield{author}{\bibinfo{person}{Wei Yang}, \bibinfo{person}{Kuang Lu},
  \bibinfo{person}{Peilin Yang}, {and} \bibinfo{person}{Jimmy Lin}.}
  \bibinfo{year}{2019}\natexlab{}.
\newblock \showarticletitle{Critically Examining the "Neural Hype": Weak
  Baselines and the Additivity of Effectiveness Gains from Neural Ranking
  Models}. In \bibinfo{booktitle}{\emph{Proceedings of the 42nd International
  {ACM} {SIGIR} Conference on Research and Development in Information
  Retrieval, {SIGIR} 2019, Paris, France, July 21-25, 2019}}.
  \bibinfo{pages}{1129--1132}.
\newblock


\bibitem[\protect\citeauthoryear{Zamani and Croft}{Zamani and Croft}{2017}]%
        {zamani2017relevance}
\bibfield{author}{\bibinfo{person}{Hamed Zamani} {and}
  \bibinfo{person}{W.~Bruce Croft}.} \bibinfo{year}{2017}\natexlab{}.
\newblock \showarticletitle{Relevance-based Word Embedding}. In
  \bibinfo{booktitle}{\emph{Proceedings of the 40th International {ACM} {SIGIR}
  Conference on Research and Development in Information Retrieval, Shinjuku,
  Tokyo, Japan, August 7-11, 2017}}. \bibinfo{pages}{505--514}.
\newblock


\bibitem[\protect\citeauthoryear{Zamani and Croft}{Zamani and Croft}{2018}]%
        {zamani2018theory}
\bibfield{author}{\bibinfo{person}{Hamed Zamani} {and}
  \bibinfo{person}{W.~Bruce Croft}.} \bibinfo{year}{2018}\natexlab{}.
\newblock \showarticletitle{On the Theory of Weak Supervision for Information
  Retrieval}. In \bibinfo{booktitle}{\emph{Proceedings of the 2018 {ACM}
  {SIGIR} International Conference on Theory of Information Retrieval, {ICTIR}
  2018, Tianjin, China, September 14-17, 2018}}. \bibinfo{pages}{147--154}.
\newblock


\bibitem[\protect\citeauthoryear{Zamani, Croft, and Culpepper}{Zamani
  et~al\mbox{.}}{2018}]%
        {zamani2018neural}
\bibfield{author}{\bibinfo{person}{Hamed Zamani}, \bibinfo{person}{W.~Bruce
  Croft}, {and} \bibinfo{person}{J.~Shane Culpepper}.}
  \bibinfo{year}{2018}\natexlab{}.
\newblock \showarticletitle{Neural Query Performance Prediction using Weak
  Supervision from Multiple Signals}. In \bibinfo{booktitle}{\emph{The 41st
  International {ACM} {SIGIR} Conference on Research {\&} Development in
  Information Retrieval, {SIGIR} 2018, Ann Arbor, MI, USA, July 08-12, 2018}}.
  \bibinfo{pages}{105--114}.
\newblock


\bibitem[\protect\citeauthoryear{Zeng, Xu, Lan, Guo, and Cheng}{Zeng
  et~al\mbox{.}}{2018}]%
        {zeng2018multi}
\bibfield{author}{\bibinfo{person}{Wei Zeng}, \bibinfo{person}{Jun Xu},
  \bibinfo{person}{Yanyan Lan}, \bibinfo{person}{Jiafeng Guo}, {and}
  \bibinfo{person}{Xueqi Cheng}.} \bibinfo{year}{2018}\natexlab{}.
\newblock \showarticletitle{Multi Page Search with Reinforcement Learning to
  Rank}. In \bibinfo{booktitle}{\emph{Proceedings of the 2018 {ACM} {SIGIR}
  International Conference on Theory of Information Retrieval, {ICTIR} 2018,
  Tianjin, China, September 14-17, 2018}}. \bibinfo{pages}{175--178}.
\newblock


\bibitem[\protect\citeauthoryear{Zhang, Song, Xiong, Rosset, Bennett, Craswell,
  and Tiwary}{Zhang et~al\mbox{.}}{2019}]%
        {zhang2019generic}
\bibfield{author}{\bibinfo{person}{Hongfei Zhang}, \bibinfo{person}{Xia Song},
  \bibinfo{person}{Chenyan Xiong}, \bibinfo{person}{Corby Rosset},
  \bibinfo{person}{Paul~N. Bennett}, \bibinfo{person}{Nick Craswell}, {and}
  \bibinfo{person}{Saurabh Tiwary}.} \bibinfo{year}{2019}\natexlab{}.
\newblock \showarticletitle{Generic Intent Representation in Web Search}. In
  \bibinfo{booktitle}{\emph{Proceedings of the 42nd International {ACM} {SIGIR}
  Conference on Research and Development in Information Retrieval, {SIGIR}
  2019, Paris, France, July 21-25, 2019}}. \bibinfo{pages}{65--74}.
\newblock


\bibitem[\protect\citeauthoryear{Zheng, Fan, Liu, Luo, Zhang, and Ma}{Zheng
  et~al\mbox{.}}{2018}]%
        {zheng2018sogou}
\bibfield{author}{\bibinfo{person}{Yukun Zheng}, \bibinfo{person}{Zhen Fan},
  \bibinfo{person}{Yiqun Liu}, \bibinfo{person}{Cheng Luo},
  \bibinfo{person}{Min Zhang}, {and} \bibinfo{person}{Shaoping Ma}.}
  \bibinfo{year}{2018}\natexlab{}.
\newblock \showarticletitle{Sogou-QCL: {A} New Dataset with Click Relevance
  Label}. In \bibinfo{booktitle}{\emph{The 41st International {ACM} {SIGIR}
  Conference on Research {\&} Development in Information Retrieval, {SIGIR}
  2018, Ann Arbor, MI, USA, July 08-12, 2018}}. \bibinfo{pages}{1117--1120}.
\newblock


\bibitem[\protect\citeauthoryear{Zheng, Liu, Fan, Luo, Ai, Zhang, and Ma}{Zheng
  et~al\mbox{.}}{2019}]%
        {zheng2019investigating}
\bibfield{author}{\bibinfo{person}{Yukun Zheng}, \bibinfo{person}{Yiqun Liu},
  \bibinfo{person}{Zhi-Qiang Fan}, \bibinfo{person}{Cheng Luo},
  \bibinfo{person}{Qingyao Ai}, \bibinfo{person}{Min Zhang}, {and}
  \bibinfo{person}{Shaoping Ma}.} \bibinfo{year}{2019}\natexlab{}.
\newblock \showarticletitle{Investigating Weak Supervision in Deep Ranking}.
\newblock \bibinfo{journal}{\emph{Data and Information Management}}
  \bibinfo{volume}{3} (\bibinfo{year}{2019}), \bibinfo{pages}{155 -- 164}.
\newblock


\end{thebibliography}
